\documentclass[reprint,onecolumn,notitlepage,showpacs,preprintnumbers,nofootinbib,amsmath,amssymb,aps,prd,floatfix,superscriptaddress,longbibliography]{revtex4-1}
\usepackage{graphicx}
\usepackage{bm}
\usepackage{subfigure}
\usepackage{url}
\usepackage[hyperindex]{hyperref}
\usepackage[svgnames]{xcolor}
\usepackage[ddmmyy,24hr]{datetime}
\usepackage{bigdelim}
\usepackage{booktabs}
\usepackage{dcolumn}
\usepackage{multirow}
\usepackage{subfigure}
\usepackage{cancel}
\usepackage{stackrel}
\usepackage{paralist}
\usepackage{xspace}
\usepackage{slashed}
\usepackage{enumerate}
\usepackage{enumitem}
\usepackage{soul}
\usepackage{lineno}

\newcommand{\nua}[1]{\ensuremath{\rlap{\kern-2.5pt\ensuremath{\overset{\scriptscriptstyle(-)}{\phantom{\nu}}}}{\ensuremath{{\nu}_{#1}}}}}

\begin{document}
	\title{Neutrinoless double beta decay in the minimal type-I seesaw model: How the enhancement or cancellation happens?}
	\author{Dong-Liang~Fang}
	\email{dlfang@impcas.ac.cn}	
	\affiliation{Institute of Modern Physics, Chinese Academy of sciences, Lanzhou, 730000, China}
	\affiliation{School of Nuclear Science and Technology, University of Chinese Academy of Sciences, Beijing 100049, China}

	\author{Yu-Feng~Li}
	\email{liyufeng@ihep.ac.cn}
	\affiliation{Institute of High Energy Physics,
		Chinese Academy of Sciences, Beijing 100049, China}
	\affiliation{School of Physical Sciences, University of Chinese Academy of Sciences, Beijing 100049, China}
	
	\author{Yi-Yu~Zhang}
	\email{zhangyiyu@ihep.ac.cn} 
	\thanks{Corresponding author}
	\affiliation{Institute of High Energy Physics,
		Chinese Academy of Sciences, Beijing 100049, China}
	\affiliation{School of Physical Sciences, University of Chinese Academy of Sciences, Beijing 100049, China}
	
	\date{\dayofweekname{\day}{\month}{\year} \ddmmyydate\today, \currenttime}
	
	\begin{abstract}
		We discuss the contribution of right-handed neutrinos (RHNs) to the effective neutrino mass of the neutrinoless double beta decay within the minimal type-I seesaw model using the intrinsic seesaw relation of neutrino mass and mixing parameters and the relative mass dependence of the nuclear matrix elements.
		In the viable parameter space, we find the possibilities of both the enhancement and cancellation to the effective neutrino mass from RHNs.
The bounds on the parameter space of the RHNs can be determined with the effective neutrino mass extracted from neutrinoless double beta decay experiments.
	\end{abstract}
	
	\maketitle
	
	\section{Introduction}\label{sec:introduction}
	
	The observation of neutrino oscillations from the atmospheric~\cite{Super-Kamiokande:1998kpq}, solar~\cite{SNO:2001kpb,SNO:2002tuh,SNO:2002hgz}, reactor~\cite{KamLAND:2002uet,DoubleChooz:2011ymz,RENO:2012mkc,DayaBay:2012fng} and accelerator~\cite{K2K:2002icj,T2K:2011ypd,NOvA:2016kwd} neutrinos is one of the most important discoveries in particle physics, which indicates that neutrinos are massive, and is currently the only evidence for physics beyond the Standard Model (SM).
	But neutrino oscillation experiments do not allow us to determine the absolute neutrino mass scale, as well as the origin of neutrino masses. 
    To determine the absolute neutrino mass scale, three complementary methods can be used.
	The first one is the cosmological observation, which probe the direct sum of three neutrino masses~\cite{BOSS:2016wmc,Planck:2019nip,Palanque-Delabrouille:2019iyz,DiValentino:2021hoh}.
	The second one is the $\beta$-decay experiments, such as the KATRIN experiment, which gives the limit on the effective neutrino mass from the spectral fine structure near the $\beta$-decay endpoint~\cite{Kraus:2004zw,Aseev:2012zz,KATRIN:2019yun}. The third is the neutrinoless double beta decay (i.e., $0\nu\beta\beta$) experiments~\cite{Andreotti:2010vj,CUORE:2015hsf,Klapdor-Kleingrothaus:2000eir,Aalseth:2000ud,GERDA:2013vls,EXO-200:2014ofj,KamLAND-Zen:2012mmx}
	that we will discuss in this work, for recent reviews see Ref.~\cite{Dolinski:2019nrj,Bilenky:2014uka,DellOro:2016tmg}.
	
	Neutrino masses are several orders of magnitude smaller than the masses of charged leptons and quarks, and may not be (or not only) of the SM Higgs origin.
	Therefore, alternative new mechanisms of the neutrino mass generation have been proposed, of which the most plausible one is the type-I seesaw mechanism~\cite{Minkowski:1977sc,Yanagida:1979as,Gell-Mann:1979vob,Mohapatra:1979ia}.
	According to this mechanism, the small neutrino masses are related to the gauge invariant masses of the SM singlets, i.e., right-handed neutrinos (RHNs), which violates the total lepton number at a mass scale much heavier than the electroweak interaction and generates the Majorana masses of massive neutrinos.
	Within such mechanism, the neutrinos are Majorana particles, and consequently, this leads to the above mentioned lepton number violating (LNV) $0\nu\beta\beta$-decay process. 
	The nature of neutrinos, that is whether the neutrino mass is of the Majorana or Dirac type, is important for our understanding of the origin for small neutrino masses.
	With the neutrino Majorana nature, more phases (namely the Majorana phases) will be included in the neutrino mixing matrix~\cite{Xing:2020ijf,Bilenky:1980cx}. Neither results from neutrino oscillation experiments, cosmology probes, nor that from $\beta$-decay experiments depend on these Majorana phases. However, from the effective neutrino mass ($m_\mathrm{eff}$) of the $0\nu\beta\beta$ decay, such information could be extracted~\cite{Xing:2015zha,Xing:2016ymd,Ge:2016tfx,Penedo:2018kpc,Cao:2019hli}.
	    
	So far, the $0\nu\beta\beta$-decay hasn't been detected yet and the lower bounds of the decay half-lives for various isotopes are obtained from different collaborations: 
	such as CUORE~\cite{CUORE:2019yfd}, EXO-200~\cite{EXO-200:2019rkq}, GERDA~\cite{GERDA:2020xhi} and KamLAND-Zen~\cite{KamLAND-Zen:2012mmx,KamLAND-Zen:2022tow}. These results help to extract $m_\mathrm{eff}$ with account of nuclear matrix elements (NMEs) from nuclear structure calculations.
	The most stringent bound on $m_\mathrm{eff}$ at present is 36-156 meV obtained from the KamLAND-Zen experiment~\cite{KamLAND-Zen:2022tow} with the uncertainties from NMEs included. 
	Future experiments will further push the limit down to 10 meV right below the inverted hierarchy (IH) region~\cite{Chen:2016qcd,LZ:2019qdm,DARWIN:2020jme,LEGEND:2021bnm}. 

	The minimal seesaw model is the simplest model of the seesaw mechanism of neutrino masses, with the inclusion of two 
	additional RHNs~\cite{King:1999mb,Frampton:2002qc,Guo:2006qa,Xing:2020ald},
	which could be important contribution to the $0\nu\beta\beta$-decay process~{\cite{Zel:1981aug,Ibarra:2010xw,Abada:2018qok,Bolton:2019pcu,Ibarra:2011xn}}.
	When the masses of RHNs are small, because of the intrinsic seesaw model, their contribution will cancel that of active neutrinos and lead to extremely small $m_\mathrm{eff}$ approaching zero, then the $0\nu\beta\beta$-decay is forbidden~\cite{Halprin:1983ez,Leung:1984vy,Blennow:2010th}.
	In contrast,
	when the masses of RHNs are large enough, they can be regarded as short-range contribution, thus the effective neutrino mass from each mass eigenstate will be inversely proportional to the neutrino mass.
However, when the masses of RHNs are between these two extreme cases, the phenomenology of the effective neutrino mass will be rather rich and complicated, which strongly depends on the intrinsic seesaw relation of neutrino mass and mixing parameters and the relative mass dependence of the NMEs.
	In this sense, considering the contribution of RHNs, even active neutrino masses are with the inverted hierarchy (IH), the possibility of non-observation of future experiments still exists due to this cancellation.
	On the other hand, if active neutrino masses are with the normal hierarchy (NH), the future experiments still have the opportunity to observe this process with the enhancement to $m_\mathrm{eff}$ from RHNs. Therefore, thorough investigation is needed to clarify the roles of the parameters of RHNs in mapping the viable parameter space.
	
	In this paper, we study the contribution of RHNs to the $0\nu\beta\beta$-decay in the minimal type-I seesaw model. 
	First, we give a brief introduction on the theoretical framework on the neutrino mass generation and the $0\nu\beta\beta$-decay in the minimal type-I seesaw model in 
	Sec.~\ref{sec:theory}. 
	Then in Sec.~\ref{sec:result} we show the numerical results taking account of the effects of RHNs on $m_\mathrm{eff}$.
	Possible parameter regions for the enhancement and cancellation of $m_\mathrm{eff}$ are identified.
	Finally, we summarize our main conclusion in Sec.\ref{sec:conclusion}.

	\section{Theoretical Framework}\label{sec:theory}
	
	The minimal type-I seesaw model is the SM extended by two RHNs $N_{\mathrm{R}I}~(\mathrm{for}~I=1,2)$, and the Lagrangian is given by~\cite{King:1999mb,Frampton:2002qc,Guo:2006qa,Xing:2020ald}
	\begin{equation}\label{key}
		\mathcal{L} = \mathcal{L}_\mathrm{SM} 
		- \left( Y_{\alpha} \overline{L_\alpha} \Tilde{\Phi} N_{\mathrm{R}} + \frac{1}{2} \overline{N_{\mathrm{R}}^c}M_{R}N_{\mathrm{R}} + \rm{h.c.} \right),
	\end{equation}
	where $L_\alpha=(\nu_{L\alpha},e_{L\alpha})^\mathrm{T}(\alpha=e,\mu,\tau)$ and $\Phi$ are the weak doublets of left-handed lepton and Higgs particles respectively, 
	$\Tilde{\Phi}=i\sigma_2\Phi^*$ where $\sigma_2$ is the second Pauli matrix,
	and $Y_{\alpha}$ and $M_{\mathrm{R}}$ are the Yukawa coupling matrix and the Majorana mass matrix of RHNs respectively.
	The seesaw mechanism works if the Dirac masses $M_\mathrm{D}=Y_{\alpha}\langle \Phi \rangle$ are much smaller than the Majorana masses of RHNs~\cite{Minkowski:1977sc,Yanagida:1979as,Gell-Mann:1979vob,Mohapatra:1979ia}.
	After spontaneous symmetry breaking, the resultant neutrino mass term becomes
    \begin{equation}
        \mathcal{L}_\mathrm{mass}=
        -\frac{1}{2} 
        \overline{\left( \nu_\mathrm{L}, N_\mathrm{R}^{c} \right)} 
        \left(
        \begin{array}{cc}
             0 & M_\mathrm{D} \\
             M_\mathrm{D}^{\mathrm{T}} & M_\mathrm{R}
        \end{array}
        \right)
        \left(
        \begin{array}{c}
             \nu_\mathrm{L}^{c}  \\
             N_\mathrm{R} 
        \end{array}
        \right)+ \rm{h.c.}\,.
    \end{equation}
The overall $5\times5$ Majorana neutrino mass matrix can be diagonalized by a unitary mixing matrix,
	\begin{equation}
        \left(
        \begin{array}{cc}
             0 & M_\mathrm{D} \\
             M_\mathrm{D}^{\mathrm{T}} & M_\mathrm{R}
        \end{array}\right)
        =
        \left(
        \begin{array}{cc}
             U & R \\
             S & V
        \end{array}\right)
        \left(
        \begin{array}{cc}
             \hat{M}_{\nu} &0 \\
             0 & \hat{M}_{\mathrm{R}}
        \end{array}\right)
        \left(
        \begin{array}{cc}
             U & R \\
             S & V
        \end{array}\right)^{\mathrm{T}}
        ,
    \end{equation}
	where $\hat{M}_{\nu}=\mathrm{Diag}\{m_1,m_2,m_3\}$ stands for the masses of the three active neutrinos and $\hat{M}_{\mathrm{R}}=\mathrm{Diag}\{M_1,M_2\}$ stands for those of RHNs,
	while $U$, $R$, $S$, $U$ are the upper-left $3\times3$, upper-right $3\times2$, lower-left $2\times3$, lower-right $2\times2$ sub-matrices of the unitary matrix respectively~\cite{Xing:2007zj,Xing:2011ur}. 
	$U$ is the mixing matrix of active neutrinos in the SM charged current interactions, named as the PMNS matrix~\cite{Maki:1962mu,Pontecorvo:1957qd}, and $R$ is that of RHNs. In the seesaw model, $U$ is not exactly unitary in the presence of $R$, which  will lead to a well-known intrinsic relation between mixing elements in the seesaw mechanism:
	\begin{equation}\label{eq:intrinsic relation}
		\sum_{i} U_{\alpha i}^2 m_i +  \sum_{I} R_{\alpha I}^2 M_I = 0.
	\end{equation}
	 The left-handed flavor neutrinos are then written as decompositions of mass eigenstates:
	\begin{equation}\label{mix6}
		\nu_{\mathrm{L} \alpha}=\sum_{i} U_{\alpha i} \nu_{i}+\sum_{I} R_{\alpha I} N_{I}^{c},
	\end{equation}
	where $N_I$ is the mass eigenstate of RHNs.
	
	Considering the $0\nu\beta\beta$-decay process in the minimal type-I seesaw model, the decay half-life is then given by~\cite{Vergados:2016hso} 
	\begin{equation}\label{key}
		\tau_{{1}/{2}} = G g_A^4 \left\lvert M_{0\nu}(0)\cdot \frac{m_{\mathrm{eff}}}{m_e} \right\rvert^2,
	\end{equation}
	where $G$ is a kinematic phase space factor~\cite{Kotila:2012zza,Stoica:2013lka}, $g_A$ is the axial vector coupling constant, $m_e$ is the mass of the electron.
	$M_{0\nu}(0)$ is the NME when the neutrino mass in the intermediate propagator is zero, which contributes the largest uncertainty~\cite{Barea:2015kwa,Simkovic:2013qiy,Menendez:2008jp,Rath:2013fma,Rodriguez:2010mn,Fang:2018tui,Faessler:2014kka}. 
	For more details on the NMEs please see the recent reviews in Ref.\cite{Engel:2016xgb,Yao:2021wst}.
	 The corresponding effective neutrino mass is: 
	\begin{equation}\label{key}
		m_{\mathrm{eff}}=m_{\mathrm{eff}}^{\nu}+m_{\mathrm{eff}}^{N},
	\end{equation}
	where the first term in the right-hand side represents the contribution from the active neutrinos:
	\begin{equation}\label{key}
		m_{\mathrm{eff}}^{\nu} = \sum_{i=1}^{3} U_{ei}^2 m_i f_\beta(m_i),
	\end{equation}
	and the contribution of RHNs is: 
	\begin{equation}\label{key}
		m_{\mathrm{eff}}^{N} = \sum_{I=1}^{2} R_{eI}^2 M_I f_\beta(M_I),
	\end{equation}
	where the normalized NME dependence is expressed as
	\begin{equation}\label{key}
		f_\beta(M) = \frac{M_{0\nu}(M)}{M_{0\nu}(0)}\,.
	\end{equation}
	In this work, only the normalized NME dependence is relevant in the effective neutrino mass, which can be empirically expressed as 
	\begin{equation}\label{key}
		f_{\beta}(M)=\frac{\left\langle p^{2}\right\rangle}{\left\langle p^{2}\right\rangle+M^{2}},
	\end{equation}
	where the parameter $\left\langle p^{2}\right\rangle$ stands for the mean Fermi momentum of nucleons in a nucleus, and the value of $\sqrt{\langle p^{2}\rangle}$ is around 200 MeV~\cite{Faessler:2014kka}. 
	{Note that recent studies suggest new short-range contribution to the $0\nu\beta\beta$-decay process~\cite{Cirigliano:2018hja}, which may give large contribution to the NMEs~\cite{Wirth:2021pij}, and also a new empirical mass-dependent relation~\cite{Dekens:2020ttz}. However, in the current discussion we restrict ourselves to the conventional calculations and will leave this new contribution to our future studies.}
	
	Since the absolute masses of three active neutrinos are much smaller than
	$\sqrt{\langle p^{2}\rangle}$,
	therefore $f_\beta(m_i)=1$. The contribution of the active neutrinos can be simplified as
	\begin{equation}\label{key}
		m_{\mathrm{eff}}^{\nu} = \sum_{i=1}^{3} U_{ei}^2 m_i.
	\end{equation}
	If we assume neutrino mixing between the active and sterile ones is small, then the contribution of the active neutrinos can be obtained as
	\begin{equation}\label{key}
		\left|m_{\mathrm{eff}}^{\nu}\right|=\bigg\vert\sum_{i=1}^{3} U_{ei}^2 m_i\bigg\vert \simeq\left|\cos^{2}{\theta_{13}} \cos^{2}{\theta_{12}}  m_{1}+\cos^{2}{\theta_{13}} \sin^{2}{\theta_{12}} e^{i \alpha_2} m_{2}+\sin^{2}{\theta_{13}} e^{i \alpha_{3}} m_{3}\right|,
	\end{equation}
	where $\theta_{ij}$ are the mixing angles of three active neutrinos in the standard parameterization, which can be obtained from the latest neutrino oscillation experiments, see the 2020 Review of Particle Physics~\cite{ParticleDataGroup:2020ssz}.
	The mass ordering of active neutrinos is not determined yet~\cite{Esteban:2020cvm}. In the minimal type-I seesaw model, the lightest neutrino mass is equal to zero, so there exist two possibilities: the NH ($m_3>m_2>m_1=0$) and the IH ($m_2>m_1>m_3=0$).
Finally $\alpha_{i}$ stand for the relative Majorana phases associated with the Majorana mass eigenstates, which are completely unknown, in what follows we will consider $0<(\alpha_2,\alpha_3)<2\pi$ for both NH and IH cases.
Following the latest oscillation experiment results~\cite{ParticleDataGroup:2020ssz}, we obtain 
\begin{equation}
1.45~\mathrm{meV}\lesssim\left|m_{\mathrm{eff}}^{\nu}\right|\lesssim3.68~\mathrm{meV}\quad({\rm NH})\quad{\rm and}\quad18.6~\mathrm{meV}\lesssim\left|m_{\mathrm{eff}}^{\nu}\right|\lesssim48.4~\mathrm{meV}\quad({\rm IH})\,,
\label{eq:meffnua}
\end{equation}
for the NH and IH cases, respectively. 

Using the intrinsic relation between the mass and mixing elements in the seesaw mechanism as shown in Eq.~(\ref{eq:intrinsic relation}), we have
	\begin{equation}\label{eq:meffv01}
		m_{\mathrm{eff}} = m_{\mathrm{eff}}^{\nu}
		-  m_{\mathrm{eff}}^{\nu}f_\beta(M_2) 
		+  R_{e1}^2 M_1 \left[f_\beta(M_1) - f_\beta(M_2)\right]\,.
	\end{equation}
	The measured half-life can only give an absolute value of $|m_{\mathrm{eff}}|$, however, $m_{\mathrm{eff}}$ actually is a complex number as the combination of various neutrino masses, mixing angles and phases. In what follows we present some detailed analyses on $m_{\mathrm{eff}}$ case by case:
    \begin{itemize}
    \item
	Degenerate masses of RHNs: $M_1= M_2=M_N$, the effective neutrino mass is given by 
	\begin{equation}\label{key}
		m_{\mathrm{eff}}= m_{\mathrm{eff}}^{\nu}[1-f_\beta(M_N)],
	\end{equation}
	which means that when the masses of RHNs are degenerated, their contribution to the effective mass is always smaller than that of the active ones $m_{\mathrm{eff}}^{\nu}$, and thus RHNs reduce $m_{\mathrm{eff}}$.
	When the degenerated mass is compatible with the active neutrino masses $m_i$, the effective mass will be strongly suppressed, approaching to zero.
	When the degenerated mass is very large 
	$M_N\gg\sqrt{\langle p^{2}\rangle}$, the contribution of RHNs is close to zero and the effective mass will become $m_{\mathrm{eff}}^{\nu}$.
    
    \item
	The masses of RHNs are both very large but not degenerated: 
	$M_1, M_2\gg\sqrt{\langle p^{2}\rangle}$, the effective neutrino mass comes completely from the active neutrino part
	\begin{equation}\label{key}
		m_{\mathrm{eff}}= m_{\mathrm{eff}}^{\nu}\,,
	\end{equation}
	which is the same as the degenerated case with $M_N\gg\sqrt{\langle p^{2}\rangle}$. In this case, the active neutrino contribution can be expressed as a function of the parameters of RHNs: 
    $m_{\mathrm{eff}}=-R_{e1}^2 M_1-R_{e2}^2 M_2$,
    which can help us to obtain strict limits on the parameters of RHNs~\cite{Xing:2009ce}.
	
	\item
	Only one of the masses of RHNs is large enough, $M_2 \gg \sqrt{\langle p^{2}\rangle}$, the effective neutrino mass is contributed by $N_{1}$ while that of $N_{2}$ is vanishing:
	\begin{equation}\label{key}
		m_{\mathrm{eff}}=m_{\mathrm{eff}}^{\nu} + R_{e1}^2 M_1 f_\beta(M_1),
	\end{equation}
	where the phenomenon on $m_{\mathrm{eff}}$ is very different from the degenerated case, but similar as the case with one sterile neutrino~\cite{Li:2011ss,Girardi:2013zra,Giunti:2015kza,Liu:2017ago,Ge:2017erv}.
	
	\item
	The masses of RHNs are both small enough compared to $\sqrt{\langle p^{2}\rangle}$. In this case, the effective neutrino mass always equals to 0 by virtue of the intrinsic relation in Eq.~(\ref{eq:intrinsic relation}), in this case the $0\nu\beta\beta$ decay is forbidden and $0\nu\beta\beta$ experiments can not help distinguish the mass hierarchy of the active neutrinos even if the sensitivity is smaller than 10 meV. One should consider other LNV processes, such as the rare meson decays~\cite{ParticleDataGroup:2020ssz} or same-sign dilepton events at colliders~\cite{Cai:2017mow}.
    \end{itemize}
Next the most complicated cases are one or two masses of RHNs are close to $\sqrt{\langle p^{2}\rangle}$, which will be numerically discussed in the next section.
	
	\section{Numerical Results}\label{sec:result}
	Although there are three phases in $m_{\mathrm{eff}}^{\nu}$, $ R_{e1}$, $R_{e2}$, only the relative phase between $R_{e1}^2$ and $m_{\mathrm{eff}}^{\nu}$ is relevant for the absolute value of the effective neutrino mass $m_{\mathrm{eff}}$, then Eq.~(\ref{eq:meffv01}) will become
	\begin{equation}
		|m_{\mathrm{eff}}| = 
		\bigg|
		|m_{\mathrm{eff}}^{\nu}|
		-  |m_{\mathrm{eff}}^{\nu}|f_\beta(M_2) 
		+  |R_{e1}^2| 
		e^{2i\delta_{14}} M_1 \left[f_\beta(M_1) - f_\beta(M_2)\right]
		\bigg|,
		\label{eq:meff5p}
	\end{equation}
	where $2\delta_{14}=\arg(R_{e1}^2)-\arg(m_{\mathrm{eff}}^{\nu})$ is the relative phase of $R_{e1}^2$ and $m_\mathrm{eff}^{\nu}$.
	To simplify our following analytical discussion, we just consider two cases with $\delta_{14}=0$ and $\delta_{14}=\pi/2$ as our representative choices. All the other cases are between these two extreme values. Therefore, the intrinsic seesaw relation will become
	\begin{equation}
	    |m_{\mathrm{eff}}^{\nu}|\pm|R_{e1}^2|M_1\pm|R_{e2}^2| M_2=0. 
	\end{equation} 
	In the case of $\delta_{14}=0$, considering the seesaw relation, we find that the phase of $R_{e2}$ can only be $\pi/2$, 
	and the intrinsic seesaw relation will become 
	\begin{equation}\label{key}
		|R_{e2}^2| M_2 -  |R_{e1}^2| M_1 = |m_{\mathrm{eff}}^{\nu}|,
	\end{equation}
	and thus the contribution of two RHNs is opposite.
	
	 In the case of $\delta_{14}=\pi/2$, the phase of $R_{e2}$ can be $0$ or $\pi/2$. 
	 This implies that the value of the phase of $R_{e2}$ cannot be uniquely determined by the seesaw relation, but also depends on the relative size of $|R_{e1}^2| M_1$ and $|m_{\mathrm{eff}}^{\nu}|$.
	 When $|R_{e1}^2| M_1>|m_{\mathrm{eff}}^{\nu}|$, the seesaw relation will become 
	\begin{equation}\label{key}
		  |R_{e1}^2| M_1-|R_{e2}^2| M_2 = |m_{\mathrm{eff}}^{\nu}|,
	\end{equation}
	where the contribution of $N_2$ will cancel a part of the contribution of $N_1$ since the individual effective neutrino mass is proportional to $|R_{ei}^2| M_i$.
	The other case is $|R_{e1}^2| M_1<|m_{\mathrm{eff}}^{\nu}|$,  which means
	\begin{equation}\label{key}
		|R_{e2}^2| M_2+|R_{e1}^2| M_1   = |m_{\mathrm{eff}}^{\nu}|,
	\end{equation}
	the contribution of $N_2$ will have the same sign as $N_1$, leading to the additive contribution of $N_1$ and $N_2$.	

With the help of these analytical discussions, in the following we shall present our results of numerical analyses. According to Eq.~(\ref{eq:meff5p}), there are in total five physical parameters in the effective neutrino mass $|m_{\mathrm{eff}}|$, i.e., $|m_{\mathrm{eff}}^{\nu}|$, $|R^{2}_{e1}|$, $\delta_{14}$, $M_1$ and $M_2$. Since we are mainly interested in those parameters related to RHNs, $|m_{\mathrm{eff}}^{\nu}|$ will be fixed at its central values in Eq.~(\ref{eq:meffnua}) for both NH and IH cases.
	
	\begin{figure}
		\centering
		\subfigure[NH, with $\delta_{14}=\pi/2$]{
			\includegraphics[width=0.4\textwidth]{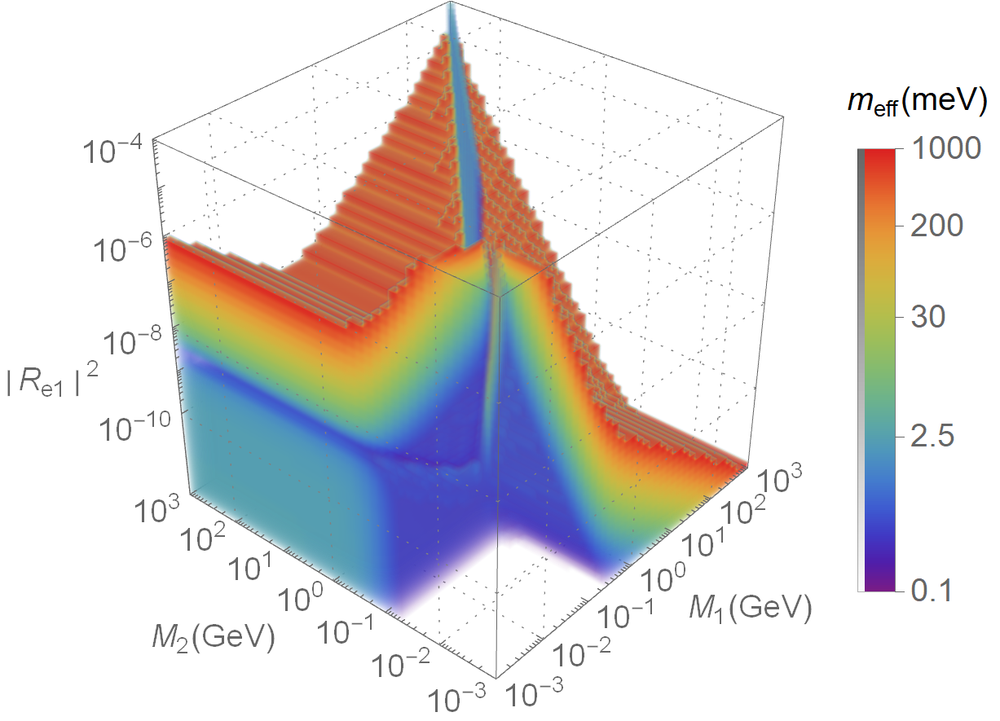}
		}
		\subfigure[NH, with $\delta_{14}=0$]{
			\includegraphics[width=0.4\textwidth]{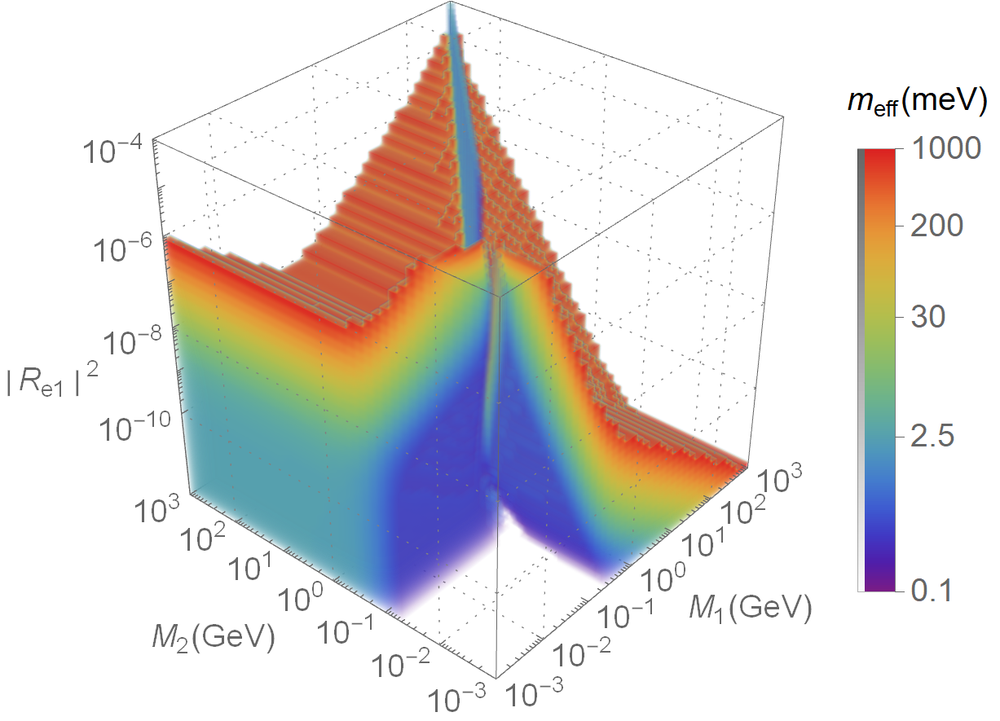}
			\label{fig:2b}
		}

		\subfigure[IH, with $\delta_{14}=\pi/2$]{
			\includegraphics[width=0.4\textwidth]{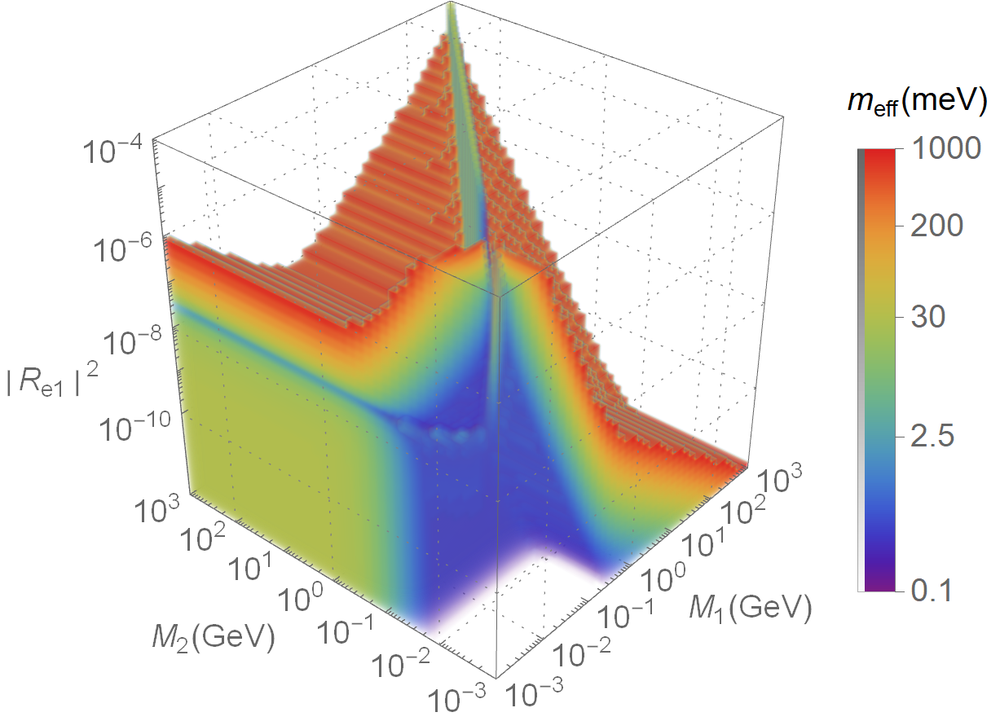}
		}
		\subfigure[IH, with $\delta_{14}=0$]{
			\includegraphics[width=0.4\textwidth]{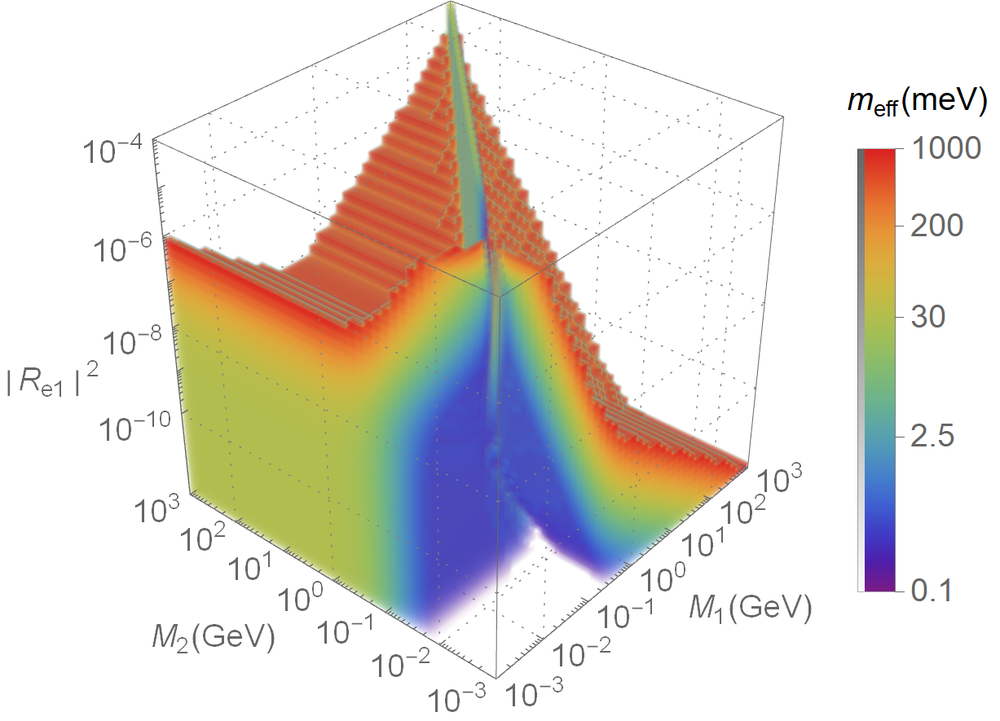}
		}
		\caption{The 4D plots of effective neutrino mass of the $0\nu\beta\beta$-decay as a function of $M_1$, $M_2$ and $|R_{e1}^2|$ for both the NH (upper panels) and IH (lower panels) cases and with both $\delta_{14}=\pi/2$ (left panels) and $\delta_{14}=0$ (right panels). The value of $m_\mathrm{eff}$ is under the logarithmic coordinates.}
		\label{fig:4D meff}
	\end{figure}
	
	\begin{figure}[htbp]
		\centering
		\subfigure[NH, with $M_2=10 ~\mathrm{MeV}$]{
			\includegraphics[width=0.35\textwidth]{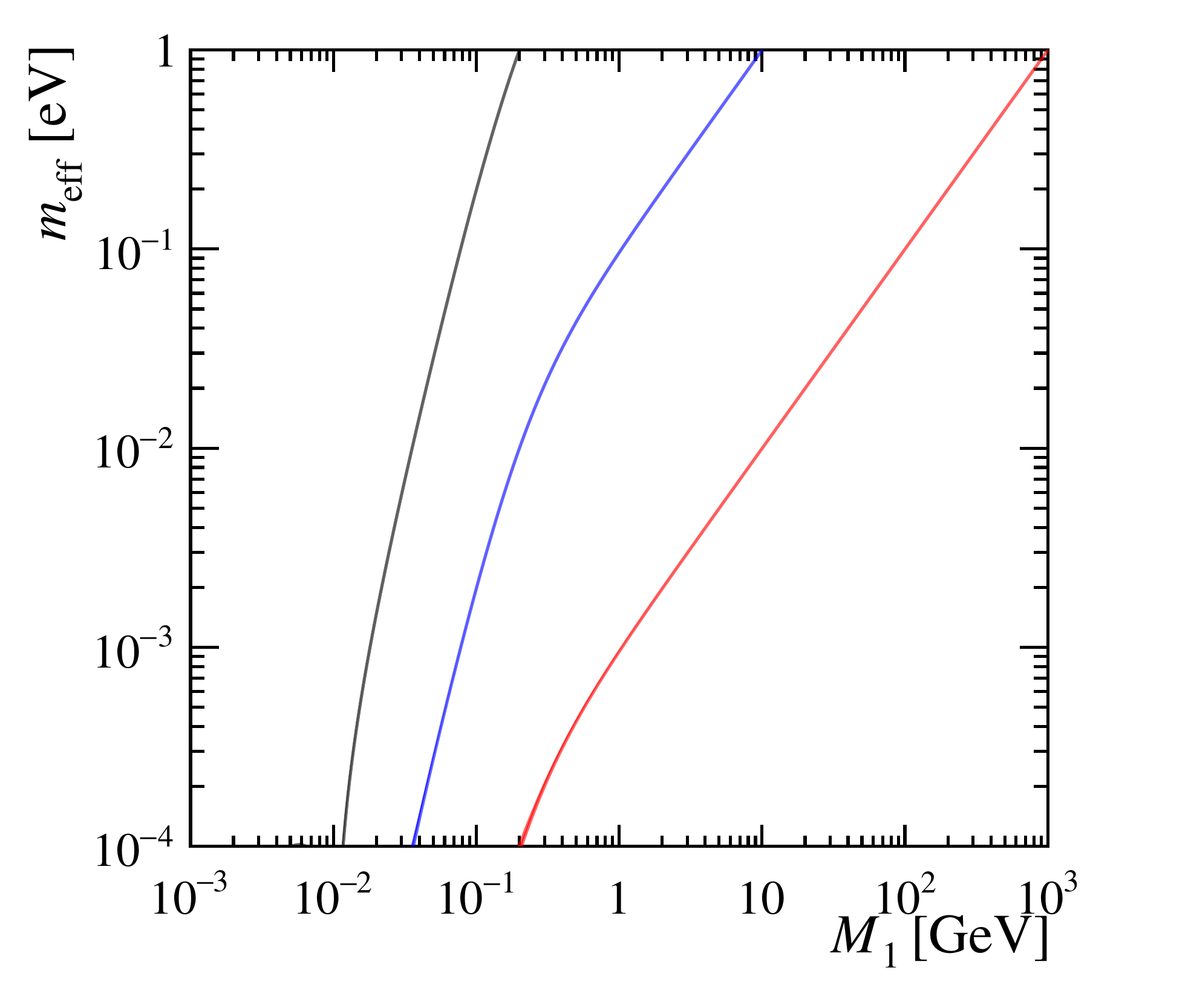}
		}
		\hspace{-1cm}
		\subfigure[NH, with $M_2=200 ~\mathrm{MeV}$]{
			\includegraphics[width=0.35\textwidth]{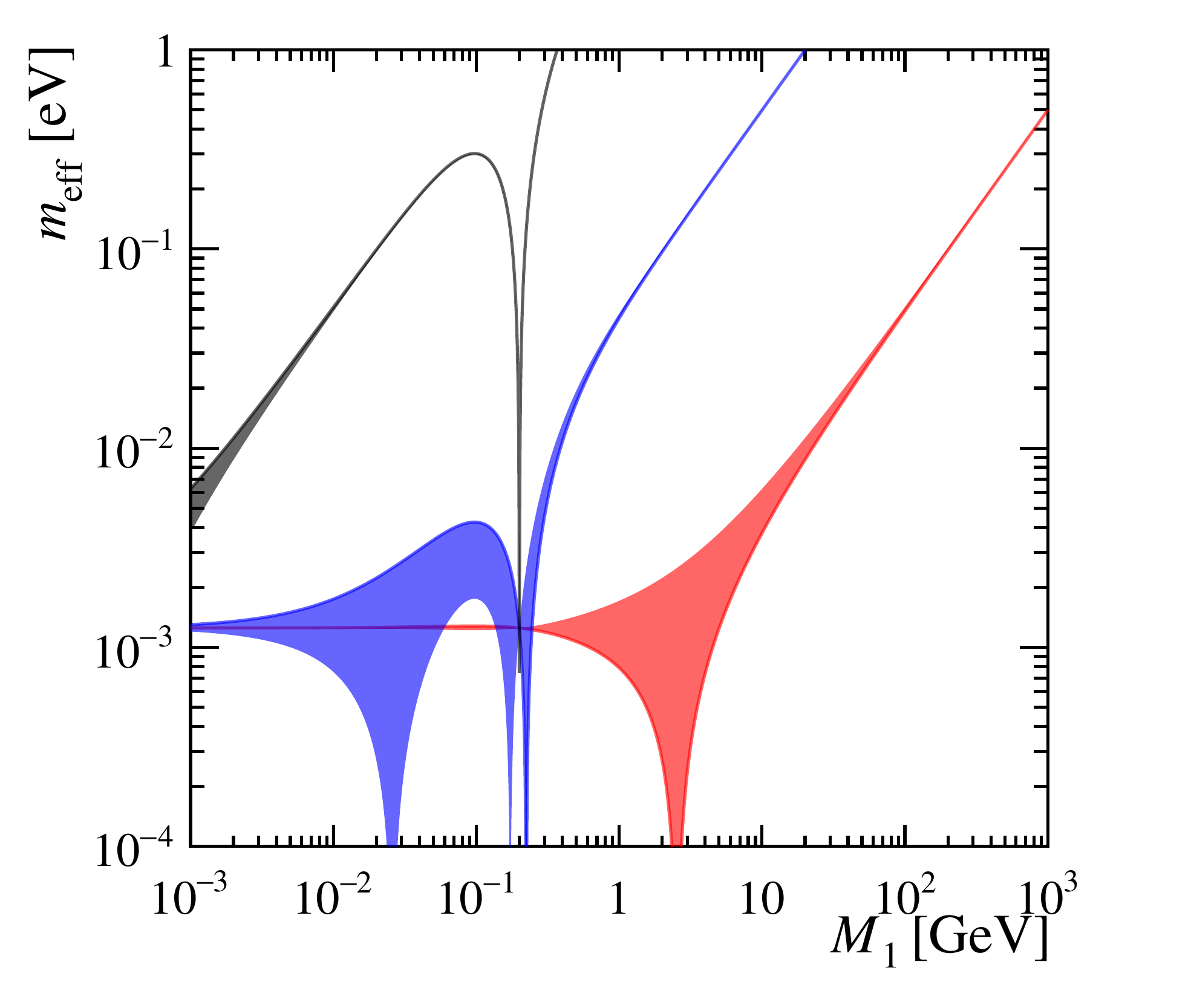}
			\label{fig:2b}
		}
		\hspace{-1cm}
		\subfigure[NH, with $M_2=1 ~\mathrm{TeV}$]{
			\includegraphics[width=0.35\textwidth]{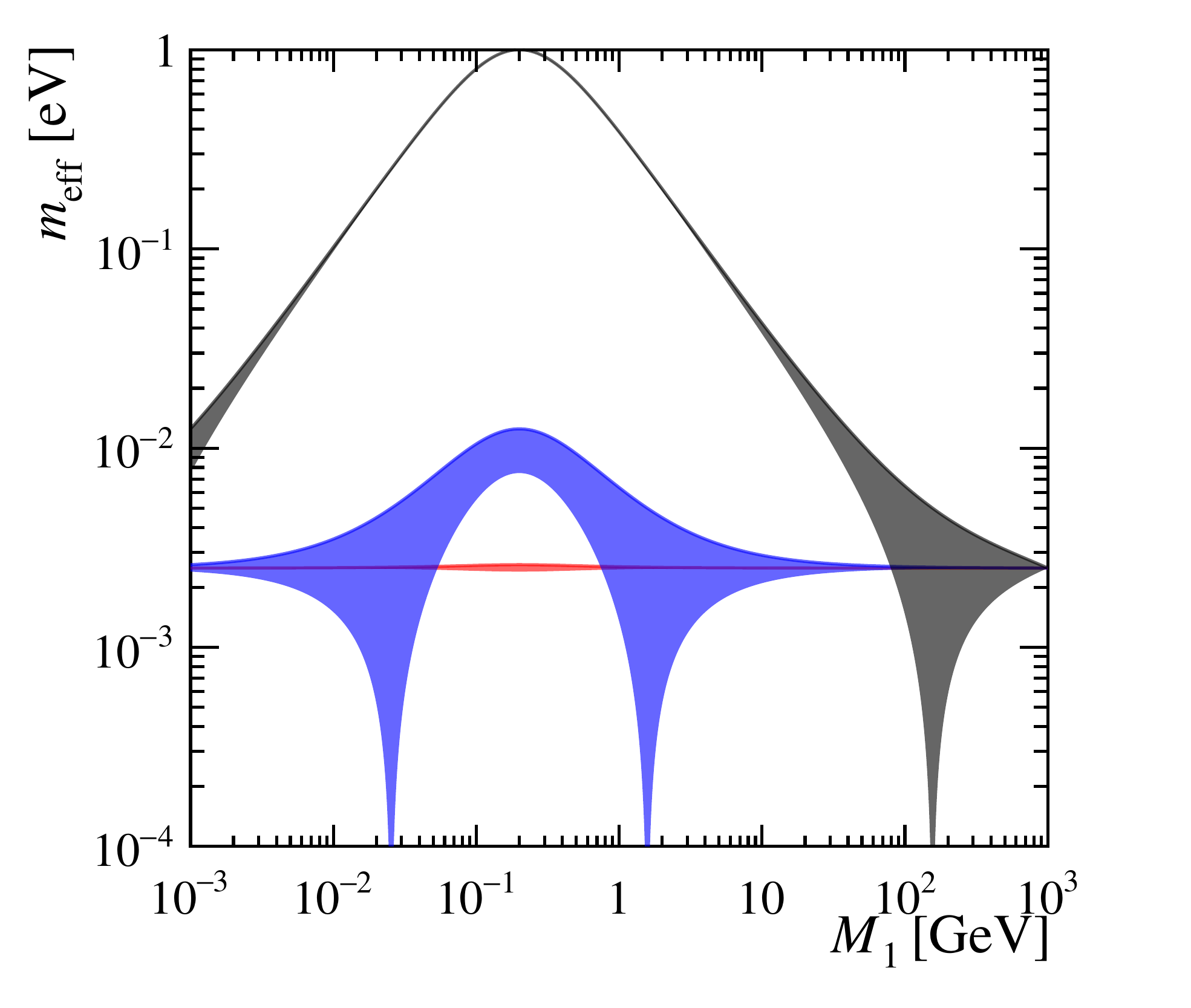}
		}
		
		\subfigure[IH, with $M_2=10 ~\mathrm{MeV}$]{
			\includegraphics[width=0.35\textwidth]{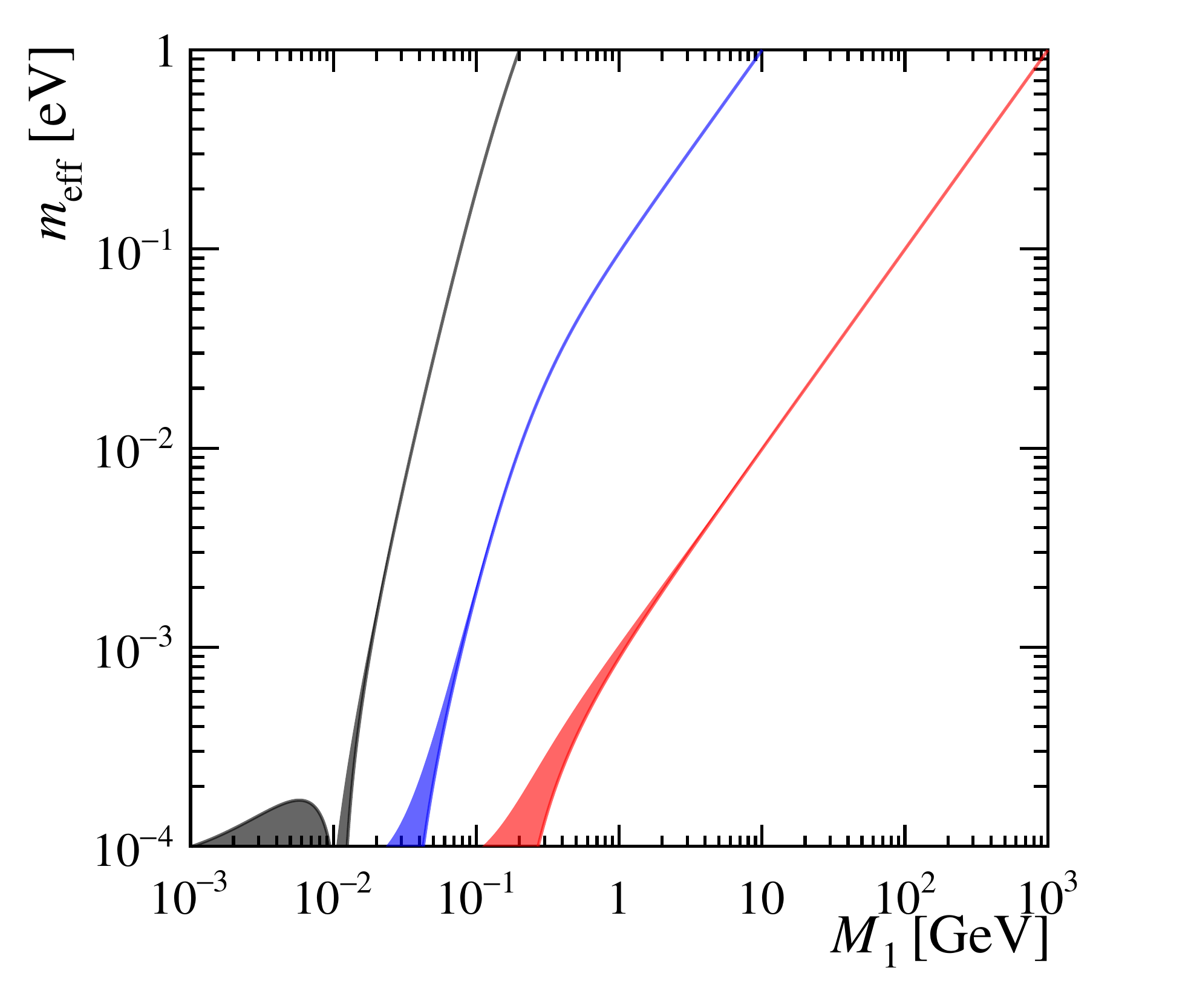}
		}
		\hspace{-1cm}
		\subfigure[IH, with $M_2=200 ~\mathrm{MeV}$]{
			\includegraphics[width=0.35\textwidth]{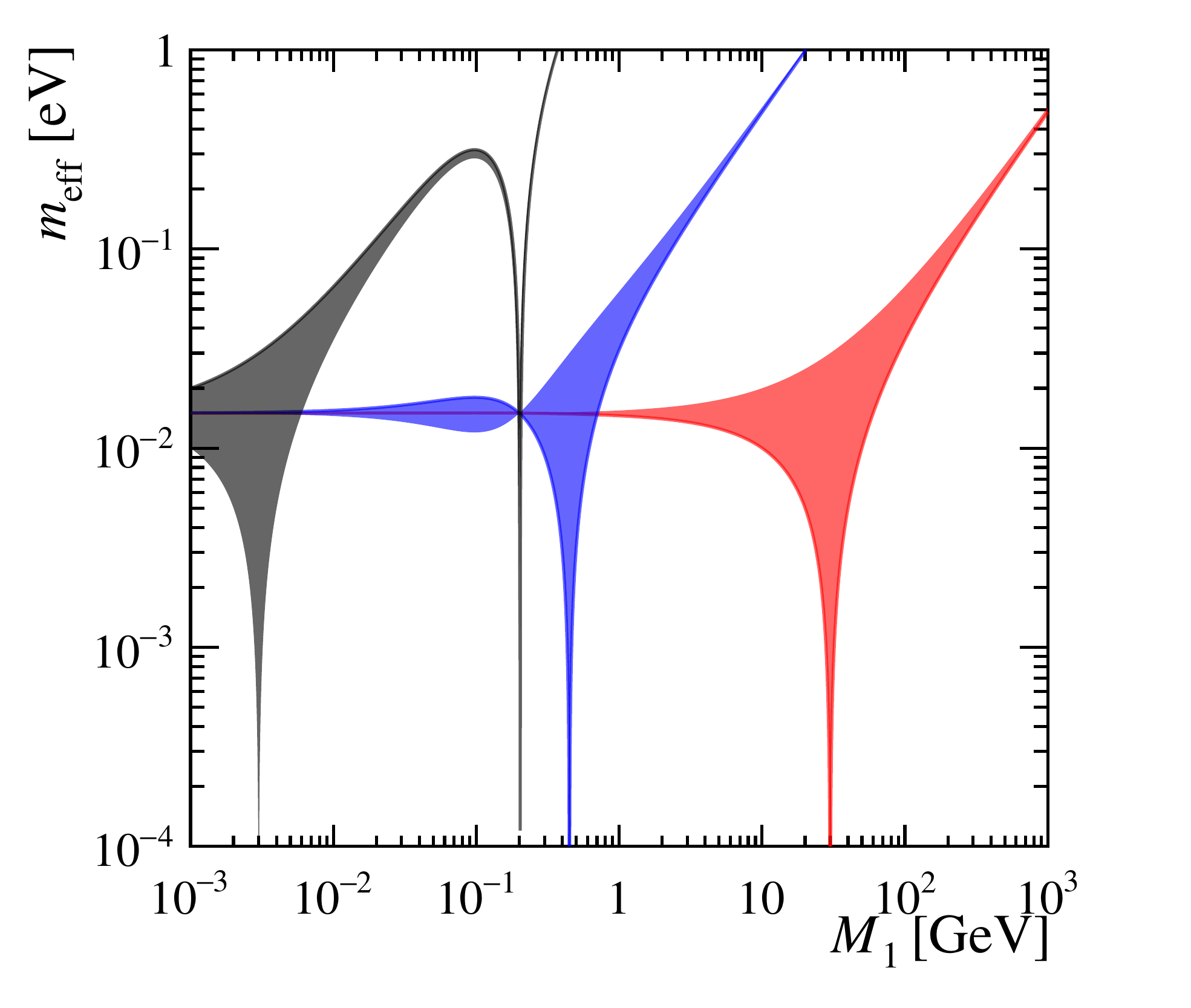}
		}
		\hspace{-1cm}
		\subfigure[IH, with $M_2=1 ~\mathrm{TeV}$]{
			\includegraphics[width=0.35\textwidth]{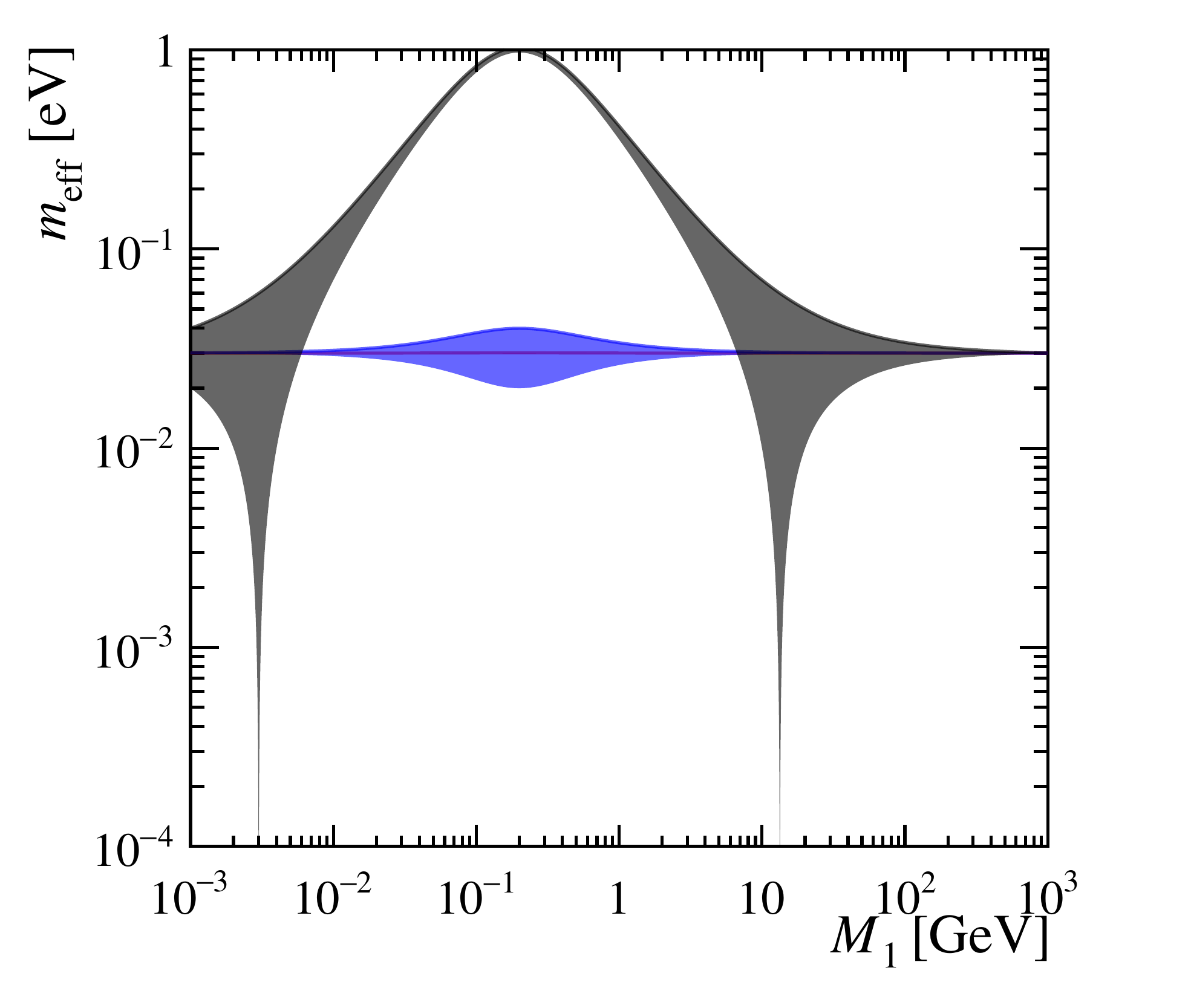}
		}
		\caption{$m_{\mathrm{eff}}$ as a function of $M_1$ for the NH (upper panels) and IH (lower panels) cases. 
		The left, middle and right panels show the cases with three different values of $M_2=0.01, 0.2, 1000 ~\mathrm{GeV}$, respectively.
		We also take $\delta_{14}$ as a free parameter with $\delta_{14}=[0,\pi/2]$ and $|R_{e1}^2|=10^{-8}$ (black), $|R_{e1}^2|=10^{-10}$ (blue), $|R_{e1}^2|=10^{-12}$ (red).}
		\label{fig:2}
	\end{figure}
	
	Fig.~\ref{fig:4D meff} shows the 4D plots of the effective neutrino mass with respect to the three free parameters $M_1$, $M_2$ and $|R_{e1}^2|$ in both the NH (upper panels) and IH (lower panels) cases and with both $\delta_{14}=\pi/2$ (left panels) and $\delta_{14}=0$ (right panels). 
	From these figures, one may roughly figure out the allowed range of the effective neutrino mass. 
	Firstly, we find that there is a plane with $M_1=M_2$ which stands for the mass degeneracy.
	The behavior for $M_1>M_2$ is very different from that of $M_1<M_2$ in both the NH and IH cases.
	In the region of $M_1<M_2$, the effective neutrino mass $m_\mathrm{eff}$ will increase as $M_2$ increases when $M_2<\sqrt{\langle p^{2}\rangle}$, and it barely changes when $M_2>1 ~\mathrm{GeV}$. Also, $m_\mathrm{eff}$ will increase as $M_1$ increases if $M_1<\sqrt{\langle p^{2}\rangle}$, and decreases otherwise. While in the region of $M_1>M_2$, $m_\mathrm{eff}$ increases as $M_1$ increases no matter $M_1$ or $M_2$ are larger or smaller than $\sqrt{\langle p^{2}\rangle}$.
	Secondly, we also find there is a special region, where the full cancellation happens. By requiring $|m_{\mathrm{eff}}|=0$, Eq.~(\ref{eq:meff5p}) becomes
	\begin{equation}\label{eq:fullconcellation}
		-\frac{|R_{e1}^2|e^{2i\delta_{14}} M_1 \left[f_\beta(M_1) - f_\beta(M_2)\right]}{1-f_\beta(M_2)}  = |m_{\mathrm{eff}}^{\nu}|.
	\end{equation}
	This suggests that the full cancellation happens only when $\delta_{14}=0$ and $\delta_{14}=\pi/2$. 
  In addition,  $M_1>M_2$ requires $\delta_{14}=0$ and when $M_1<M_2$, one should take $\delta_{14}=\pi/2$.
    If $M_1<M_2$ (thus $\delta_{14}=\pi/2$) and $M_2>1 ~\mathrm{GeV}$, the value of $f_\beta(M_2)$ is close to 0, and there are always full cancellation lines for certain values of $M_1$: $|R_{e1}^2|=|m_{\mathrm{eff}}^{\nu}|/[M_1 f_\beta(M_1)]$.
    For example, when $M_1=1~\mathrm{MeV}$, $|R_{e1}^2|$ is around $10^{-9}$ in NH (upper-left panel) and $10^{-7}$ in IH (lower-left panel).
    The similar discussion can also be found in Ref.~\cite{Asaka:2020wfo}.
       
	To uncover more detailed dependence of $m_\mathrm{eff}$ on the RHNs parameters,
	we want to fix some parameters and analyze the variations of others.
	Fig.~\ref{fig:2} shows the effective neutrino mass as a function of $M_1$ with fixed $M_2$ and $|R_{e1}^2|$ for the cases of NH (upper panels) and IH (lower panels). 
	The left, middle and right panels show the cases with three different values of $M_2=0.01, 0.2, 1000 ~\mathrm{GeV}$, respectively. The shadow regions illustrate the effect of the phase variation with $\delta_{14}=[0,\pi/2]$.
	In each plot, we also take three absolute values of the mixing matrix element $|R_{e1}^2|=10^{-8}, 10^{-10}, 10^{-12}$.
	Comparing these plots from the horizontal lines, we observe that there is a clear boundary at $M_1=M_2$ for the case of $M_2=200 ~\mathrm{MeV}$ (middle panels), and $m_{\mathrm{eff}}$ evolves differently above and below this boundary. When $M_1<M_2$, it is similar to the case of $M_2=10 ~\mathrm{MeV}$, but when $M_1>M_2$, it is similar to the case of $M_2=1 ~\mathrm{TeV}$.
	In the case of $ M_2=10 ~\mathrm{MeV}$ (left panels), a tiny mass of $M_1$ will make $m_{\mathrm{eff}}$ close to zero, as we have already discussed. Then $m_{\mathrm{eff}}$ grows as $M_1$ and $|R_{e1}^2|$ increase.
	In the case of $ M_2=1 ~\mathrm{TeV}$ (right panels), at the two extreme ends of $M_1$, $m_{\mathrm{eff}}$ is close to contribution of the active neutrinos $m_{\mathrm{eff}}^{\nu}$, which means the disappearance of the contribution from RHNs.
	When $M_1$ is around $\langle p^{2}\rangle$ with large $|R_{e1}^2|$, the contribution of RHNs increases significantly which indicates that the enhancement happens.
	We also find the regions with the mass-degeneracy effect, where $m_{\mathrm{eff}}$ will quickly converge to the value of $m_{\mathrm{eff}}^{\nu}[1-f_\beta(M_N)]$ 
	in the vicinity of $M_1=M_2$.
	There are also regions of full cancellation, when the condition of Eq.~(\ref{eq:fullconcellation}) is fulfilled.
	In these cases, the phase $\delta_{14}$ will become crucial on the effective neutrino mass around the points of $m_{\mathrm{eff}}\rightarrow 0$.
	
	\begin{figure}[!htbp]
		\centering
		\subfigure[NH, with $|m_{\mathrm{eff}}|<|m_{\mathrm{eff}}^{\nu}|$]{
			\includegraphics[width=0.4\textwidth]{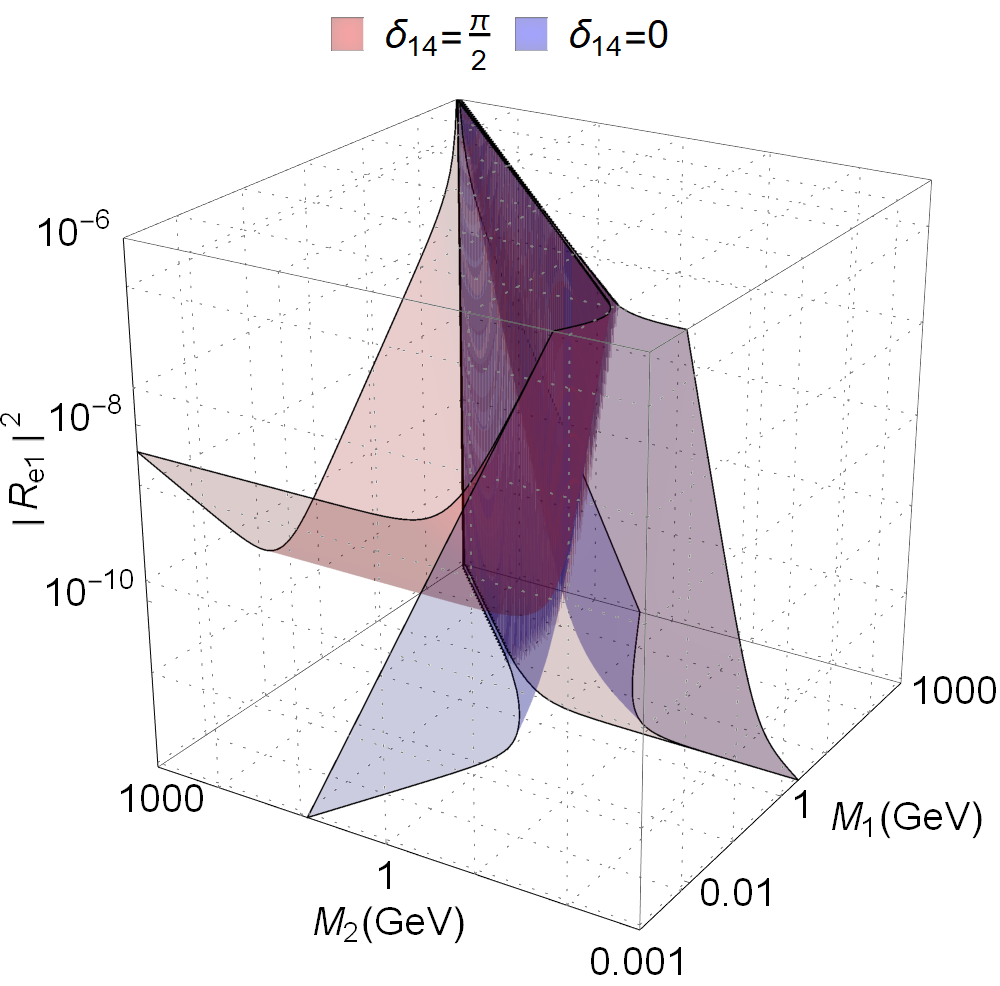}
			\label{fig:4a}
		}
		\subfigure[NH, with $|m_{\mathrm{eff}}|>200 ~\mathrm{meV}$]{
			\includegraphics[width=0.4\textwidth]{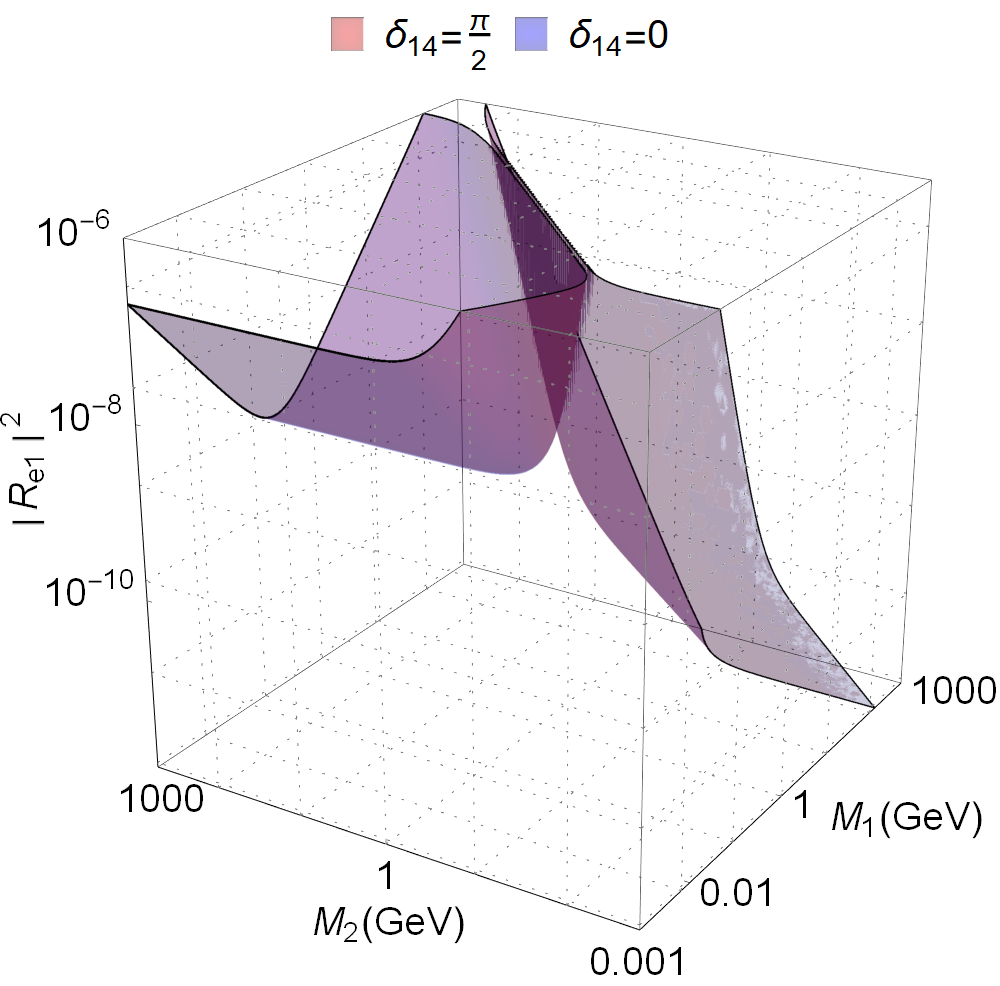}
			\label{fig:4b}
		}
		
		\subfigure[IH, with $|m_{\mathrm{eff}}|<|m_{\mathrm{eff}}^{\nu}|$]{
			\includegraphics[width=0.4\textwidth]{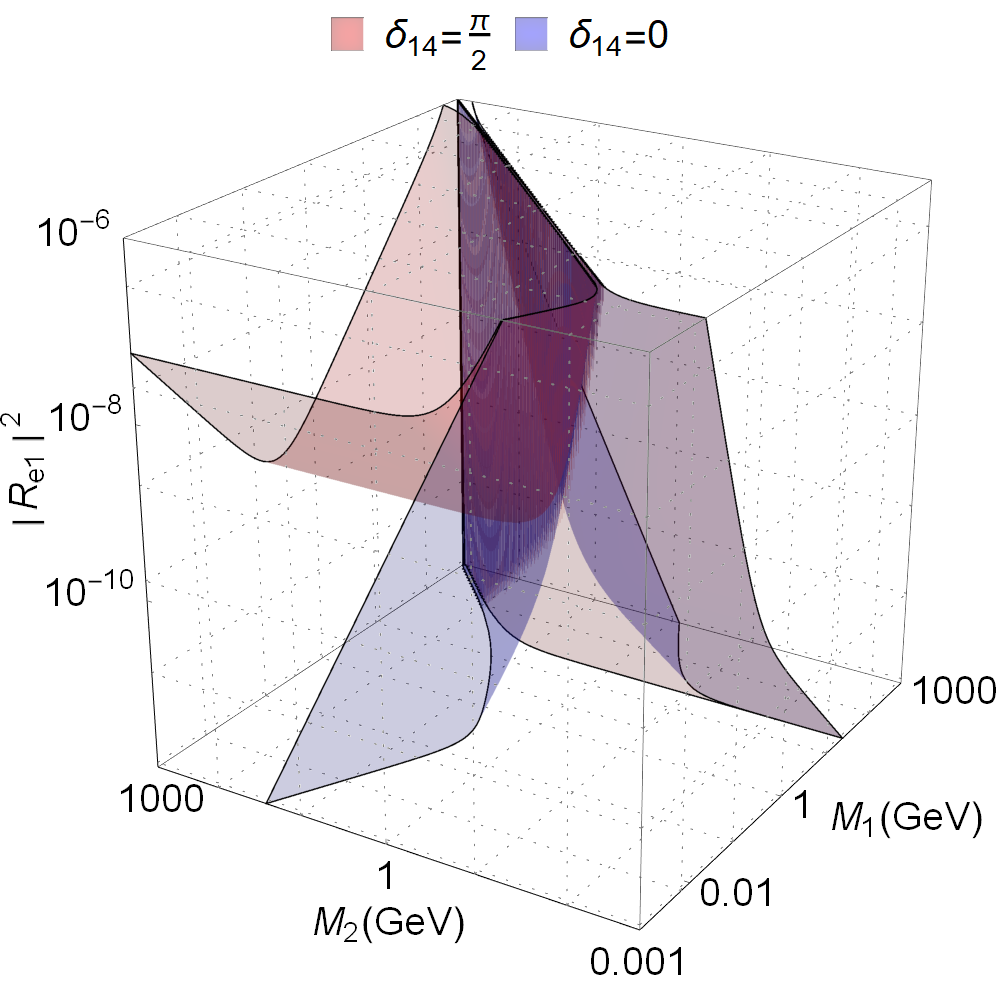}
			\label{fig:4d}
		}
		\subfigure[IH, with $|m_{\mathrm{eff}}|>200 ~\mathrm{meV}$]{
			\includegraphics[width=0.4\textwidth]{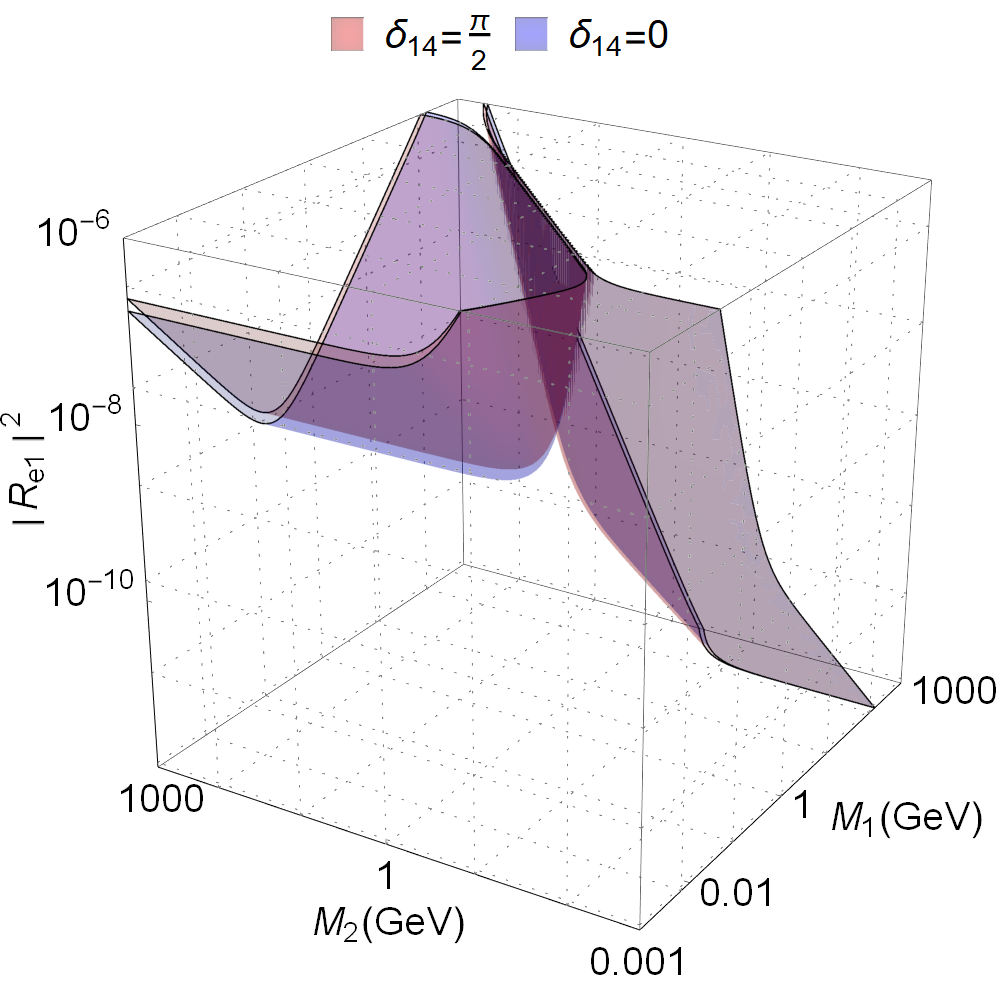}
		}
		\caption{Contours of the cancellation (left panels) and strong enhancement (right panels) regions of $m_{\mathrm{eff}}$  for the NH (upper panels) and IH (lower panels) cases. In each plot, we also take two cases with $\delta_{14}=\pi/2$ and $\delta_{14}=0$.}
		\label{fig:3d can}
	\end{figure}
	
	Next we are going to take a closer look at the properties of enhancement and cancellation. To be more specific, the cancellation is defined as the region with $|m_{\mathrm{eff}}|<|m_{\mathrm{eff}}^{\nu}|$, and the strong enhancement is defined as the region with $|m_{\mathrm{eff}}|>{\rm 200\;meV}$, which is roughly the region excluded by the current experiments~\cite{DellOro:2016tmg}.
	Between these two parts, there is a region of moderate enhancement
	with $|m_{\mathrm{eff}}^{\nu}|<|m_{\mathrm{eff}}|<{\rm 200\;meV}$.
	For this purpose, in Fig.~\ref{fig:3d can} we present the 3D contour plots of the cancellation (left panels) and strong enhancement (right panels) regions of $m_{\mathrm{eff}}$ for the NH (upper panels) and IH (lower panels) cases. In each plot, we also take two special cases with $\delta_{14}=\pi/2$ and $\delta_{14}=0$. From the figure, we can roughly see the correlations among $M_1$, $M_2$ and $|R_{e1}^2|$. 
	In general, with decreasing $|R_{e1}^2|$, the cancellation ranges will be expanded.
	The plots are divided into two different regions by the boundary $M_1=M_2$. In the region of $M_1<M_2$, the behavior of the parameters is very different for the two cases with $\delta_{14}=\pi/2$ and $\delta_{14}=0$. There is an obvious hollow around $M_1=\sqrt{\langle p^{2}\rangle}$ for $\delta_{14}=\pi/2$, but it disappears with $\delta_{14}=0$.
	In the region of $M_1>M_2$, the behavior is similar between $\delta_{14}=\pi/2$ and $\delta_{14}=0$. We also observe that the shape difference of the cancellation and strong enhancement contours is very small between the NH and IH cases.
	
	\begin{figure}[!htbp]
		\centering
		\subfigure[NH, with $|R_{e1}^2|=10^{-8}$]{
			\includegraphics[width=0.34\textwidth]{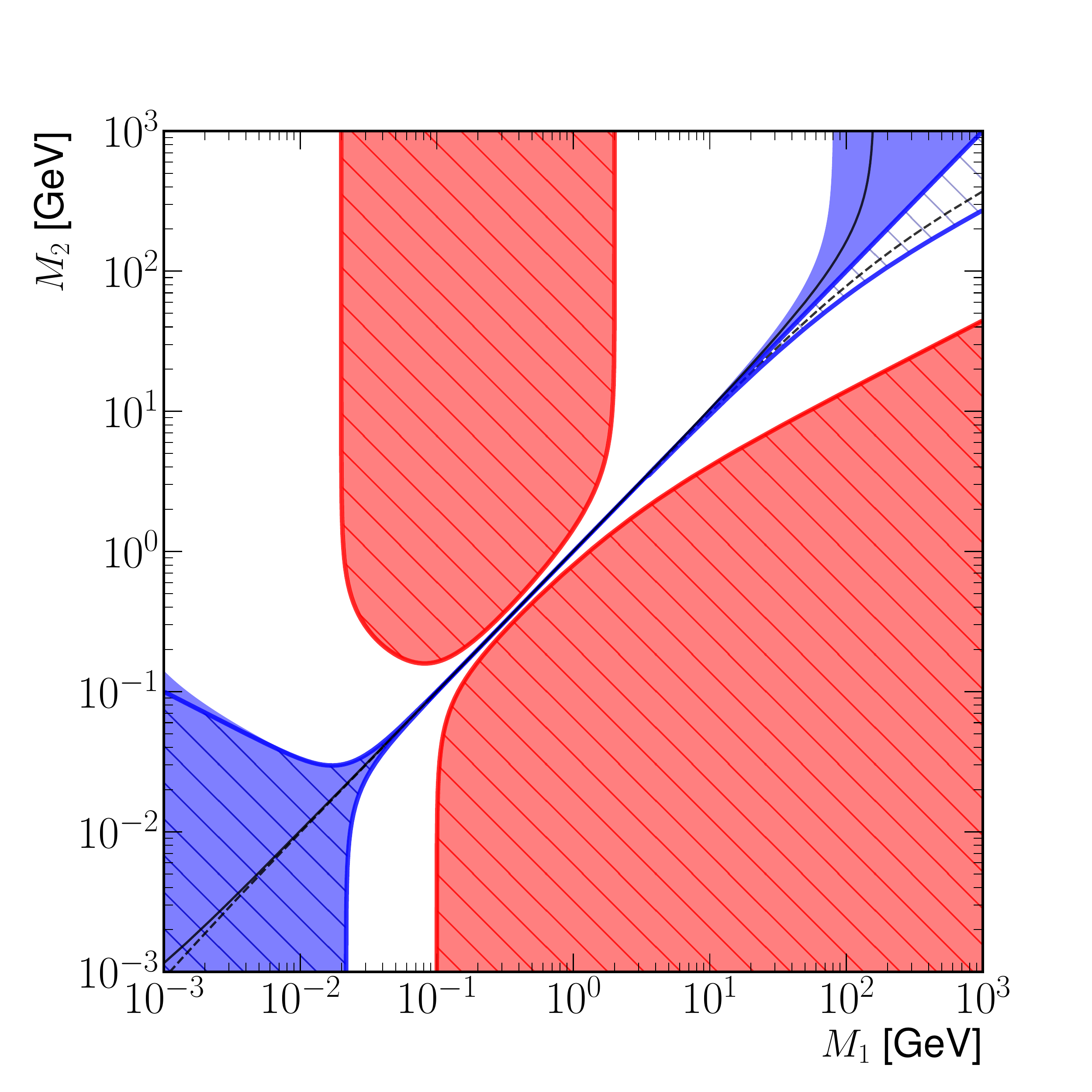}
			\label{fig:4a}
		}
		\hspace{-0.8cm}
		\subfigure[NH, with $|R_{e1}^2|=10^{-10}$]{
			\includegraphics[width=0.34\textwidth]{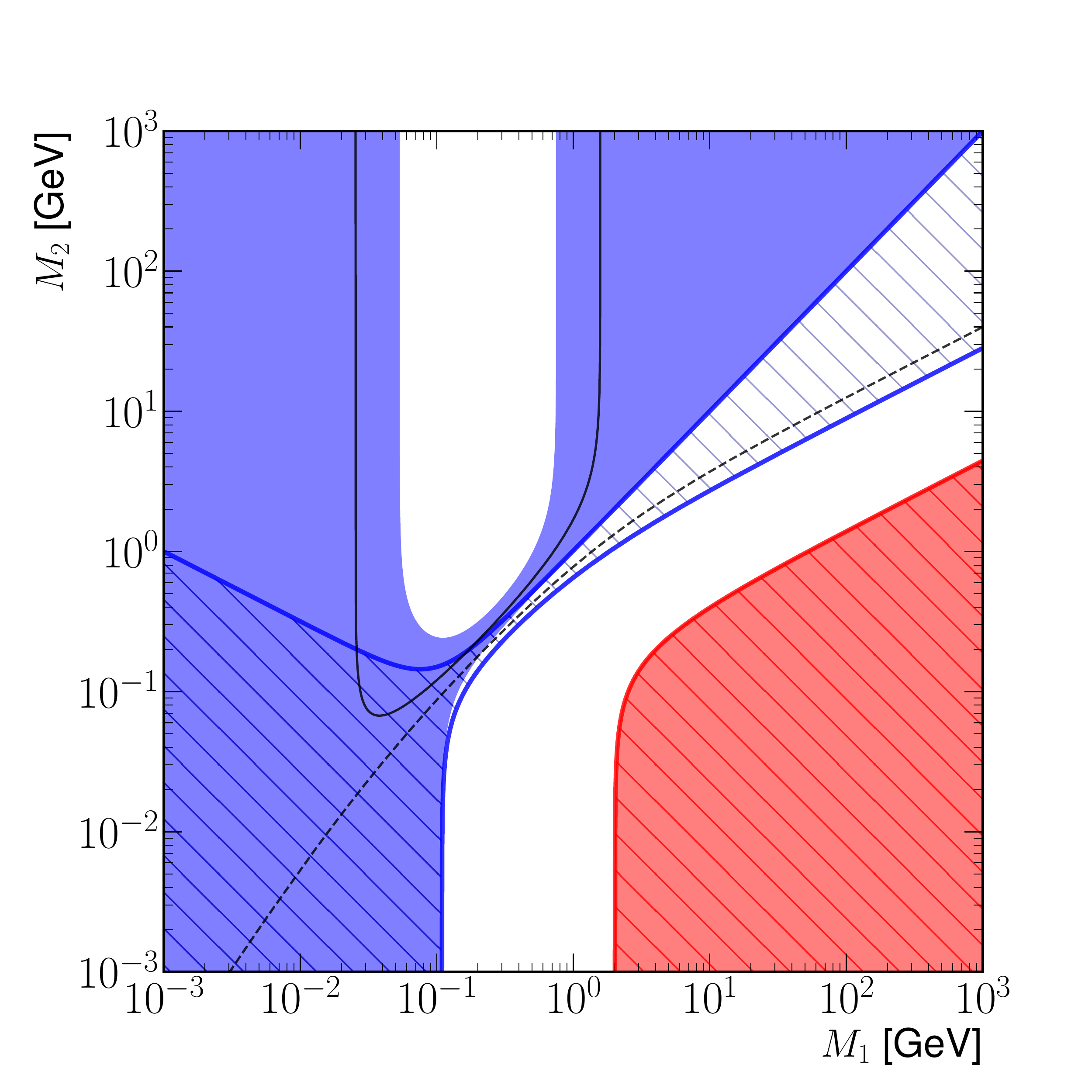}
			\label{fig:4b}
		}
		\hspace{-0.8cm}
		\subfigure[NH, with $|R_{e1}^2|=10^{-12}$]{
			\includegraphics[width=0.34\textwidth]{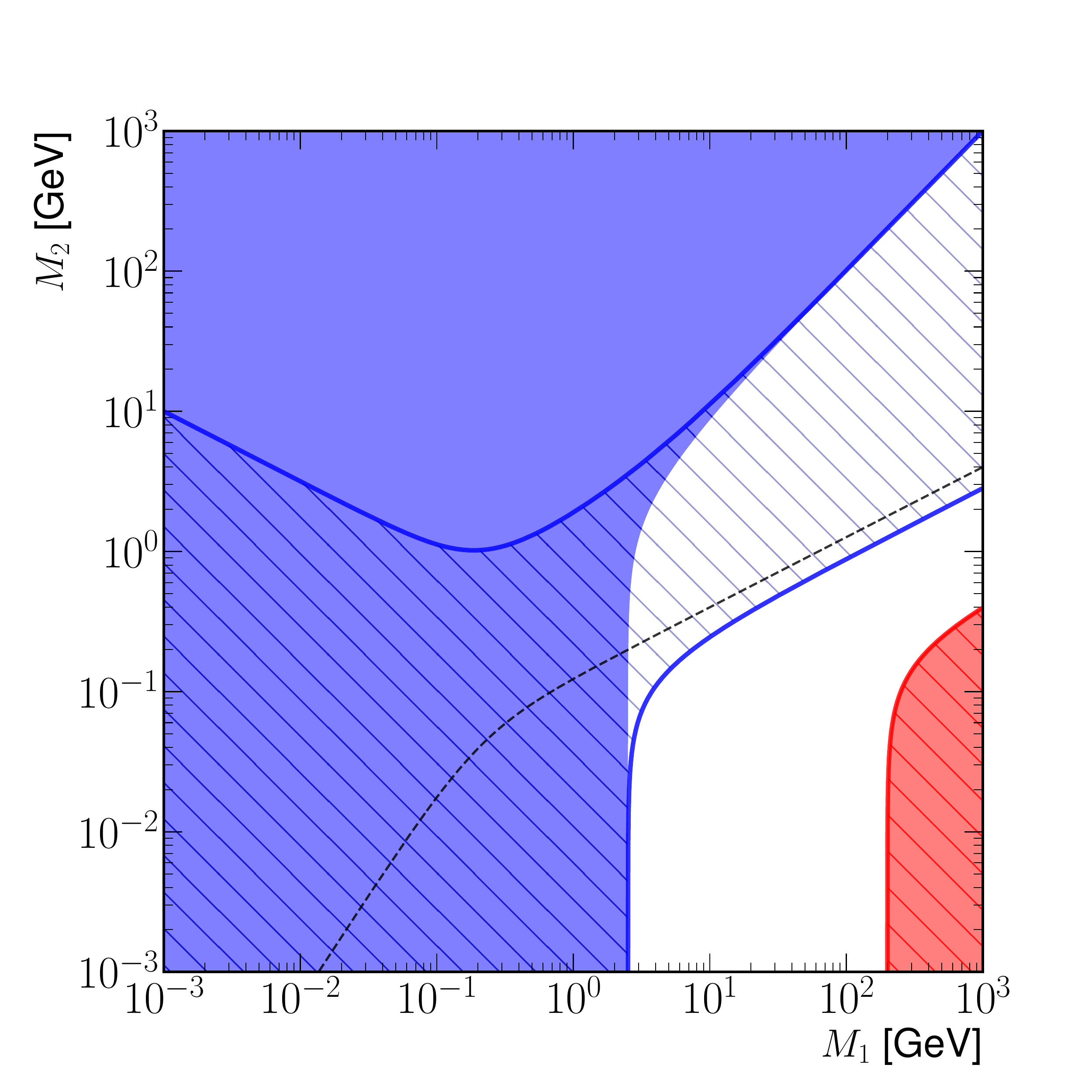}
		}
		
		\subfigure[IH, with $|R_{e1}^2|=10^{-8}$]{
			\includegraphics[width=0.34\textwidth]{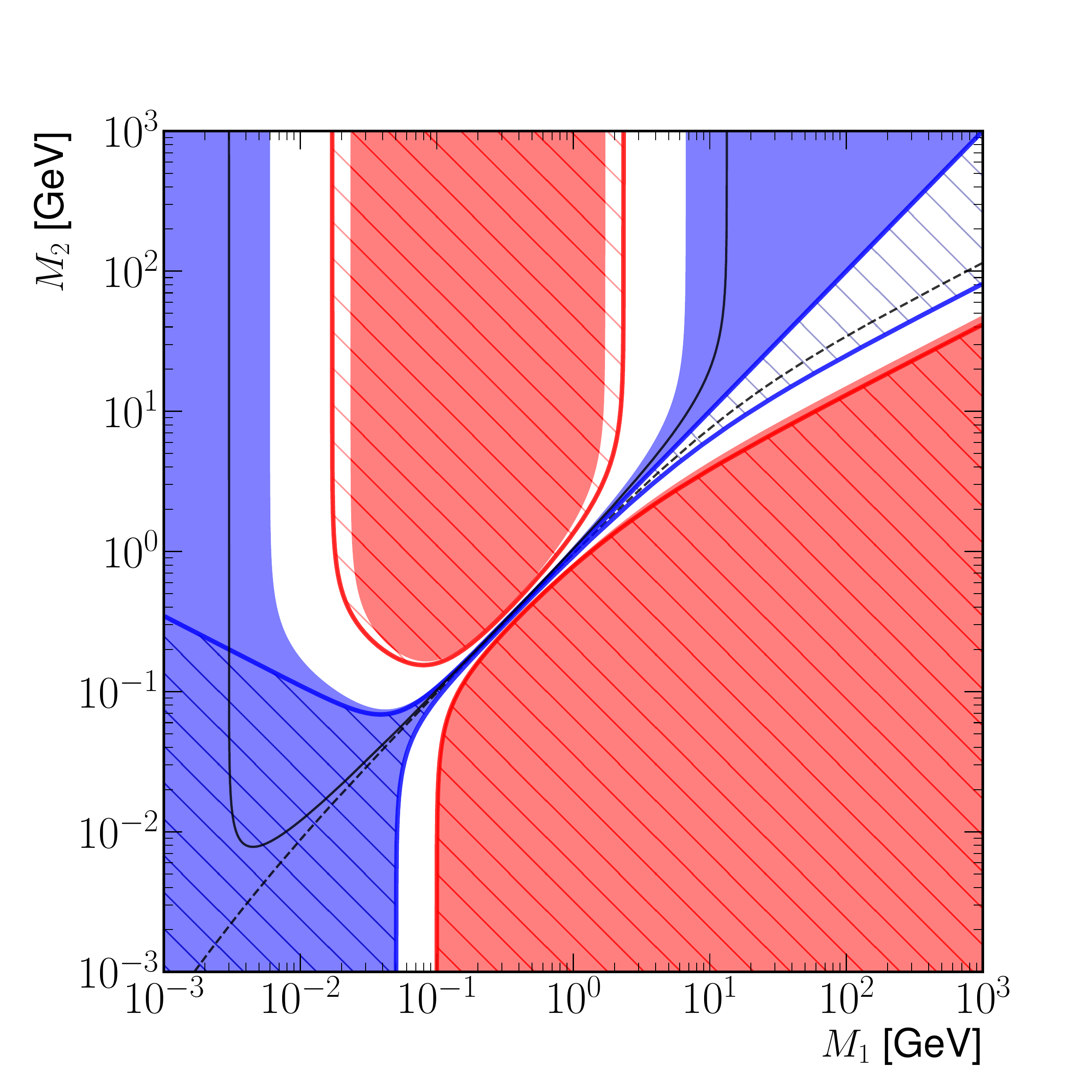}
			\label{fig:4d}
		}
		\hspace{-0.8cm}
		\subfigure[IH, with $|R_{e1}^2|=10^{-10}$]{
			\includegraphics[width=0.34\textwidth]{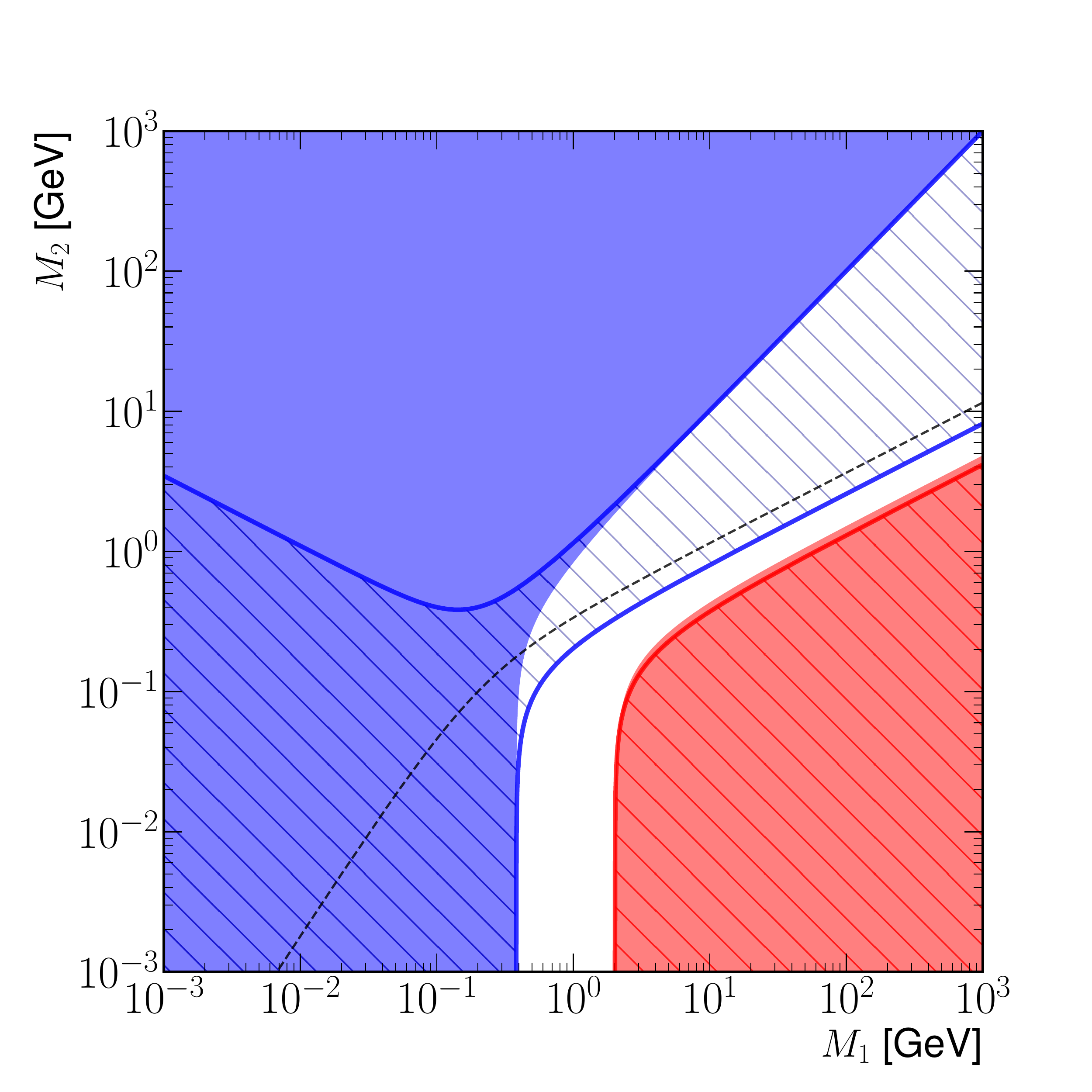}
		}
		\hspace{-0.8cm}
		\subfigure[IH, with $|R_{e1}^2|=10^{-12}$]{
			\includegraphics[width=0.34\textwidth]{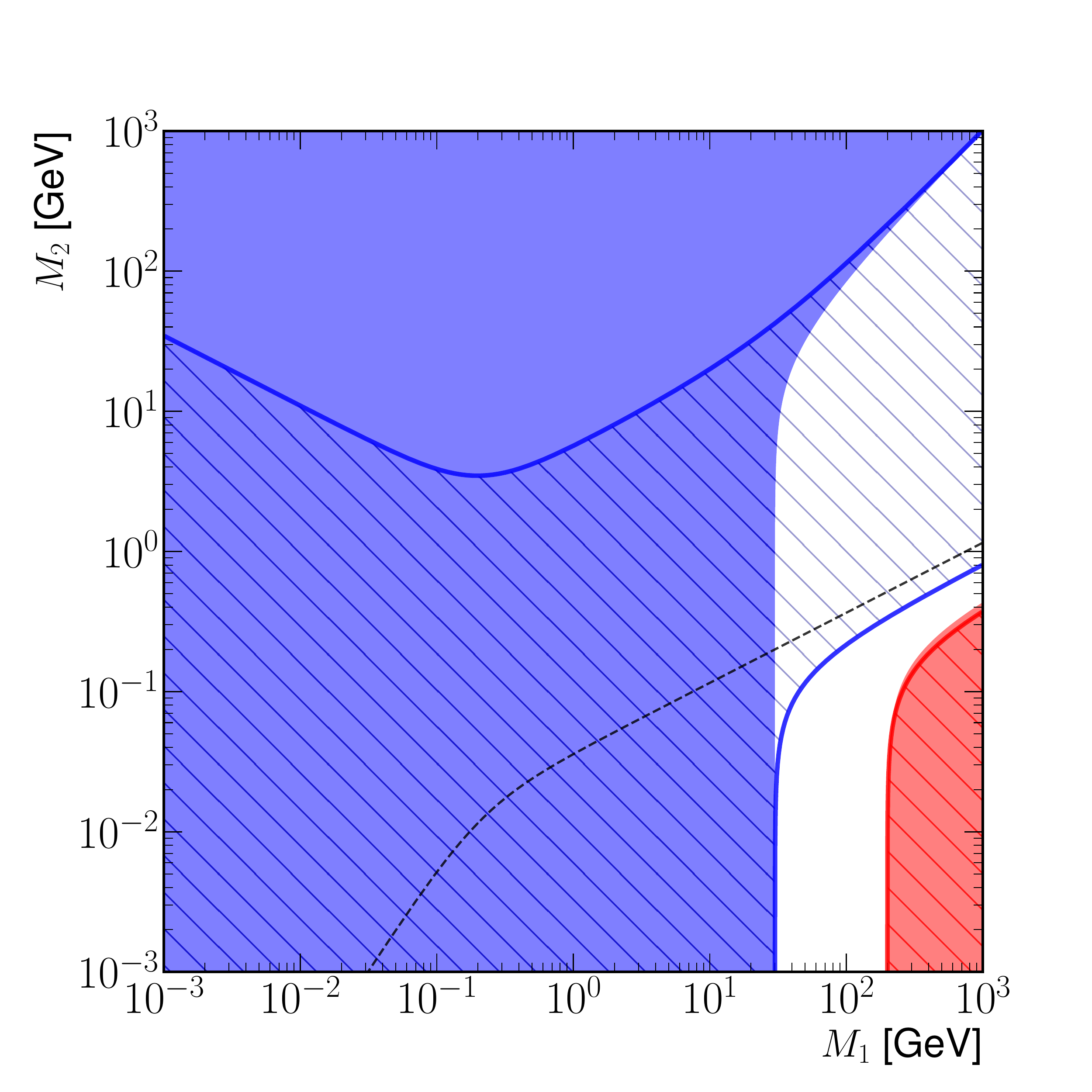}
		}
		\caption{Contours of $|m_{\mathrm{eff}}|$ in the $0\nu\beta\beta$-decay as functions of $M_{1}$ and {$M_{2}$} for the NH (top) and IH (bottom) cases with two specific values of $\delta_{14}$ with $\delta_{14}=\pi/2$ (colored regions without slash lines) and $\delta_{14}=0$ (regions with slash lines).	The black solid line and dashed line represent the special $m_{\mathrm{eff}}=0$ cases for $\delta_{14}=\pi/2$ and $\delta_{14}=0$, respectively.
		The blue regions (either light or dark) stand for $|m_{\mathrm{eff}}|<|m_{\mathrm{eff}}^{\nu}|$,
		the red regions stand for $|m_{\mathrm{eff}}|>200~\mathrm{meV}$,
		and the white regions correspond to the moderate enhancement in $|m_{\mathrm{eff}}|$. }
		\label{fig:4}
	\end{figure}

	\begin{figure}[!htbp]
		\centering
		\subfigure[NH, $M_{2}=10 ~\mathrm{MeV}$]{
			\includegraphics[width=0.34\textwidth]{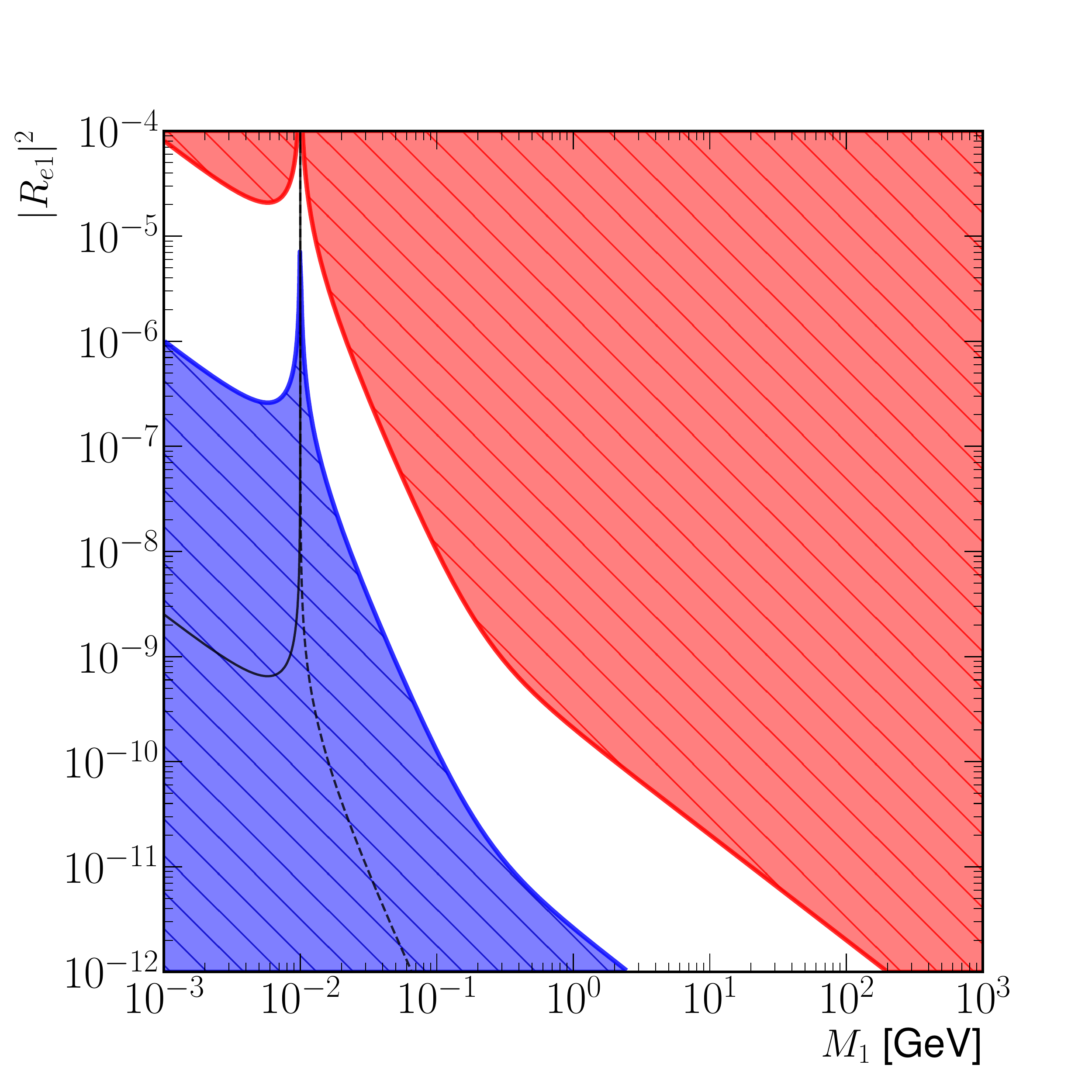}
		}
		\hspace{-0.8cm}
		\subfigure[NH, $M_{2}=200 ~\mathrm{MeV}$]{
			\includegraphics[width=0.34\textwidth]{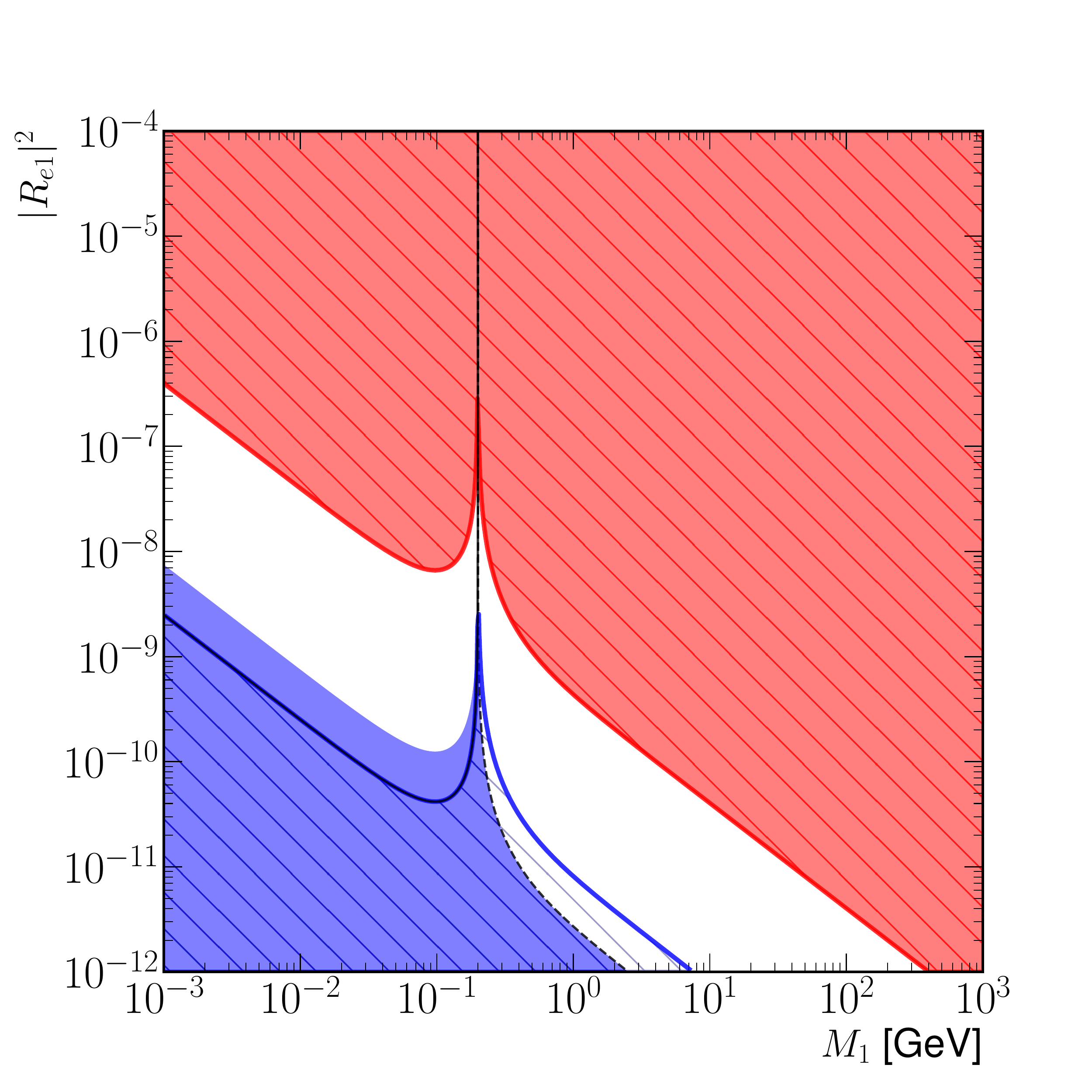}
		}
		\hspace{-0.8cm}
		\subfigure[NH, $M_{2}=1 ~\mathrm{TeV}$]{
			\includegraphics[width=0.34\textwidth]{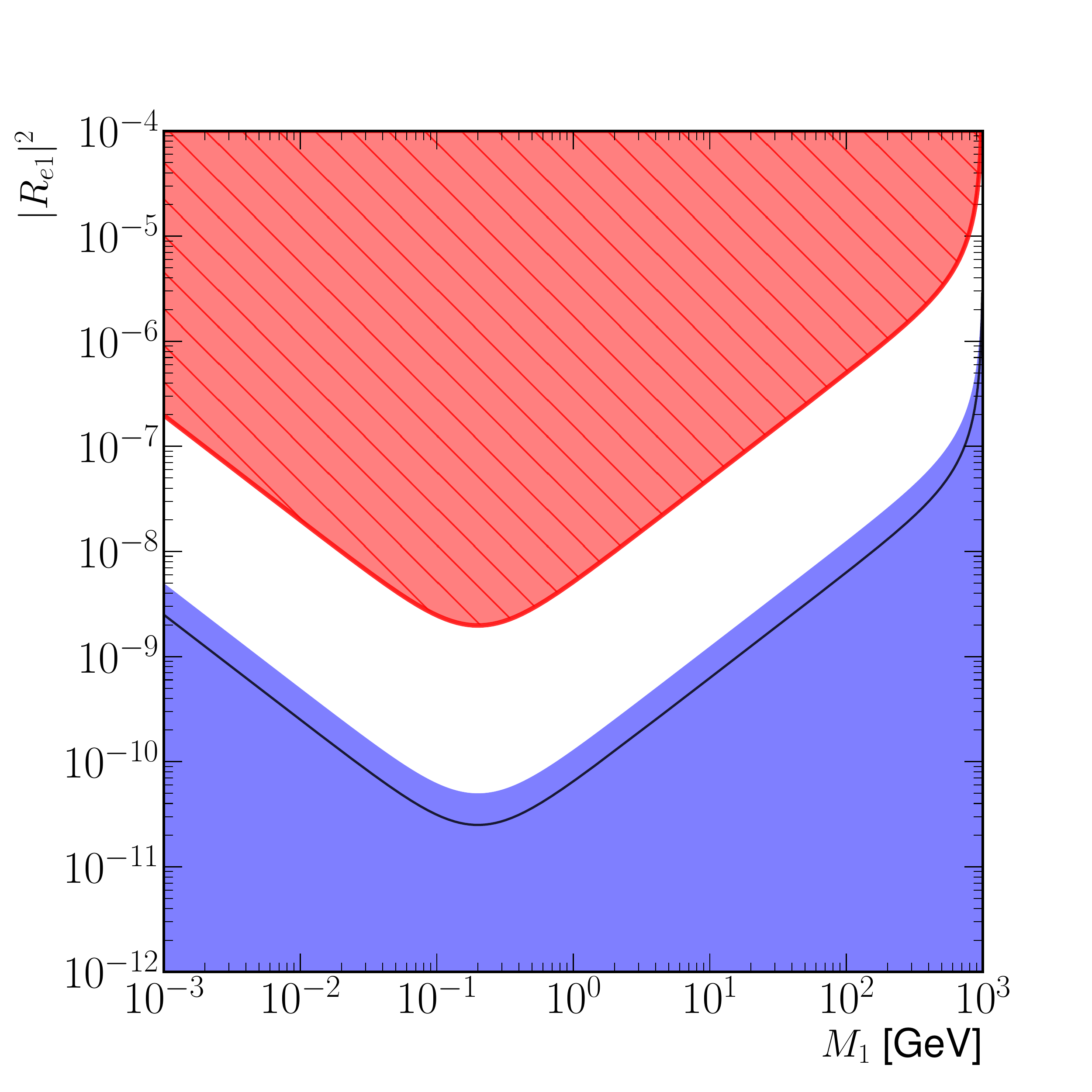}
		}
		
		\subfigure[IH, $M_{2}=10 ~\mathrm{MeV}$]{
			\includegraphics[width=0.34\textwidth]{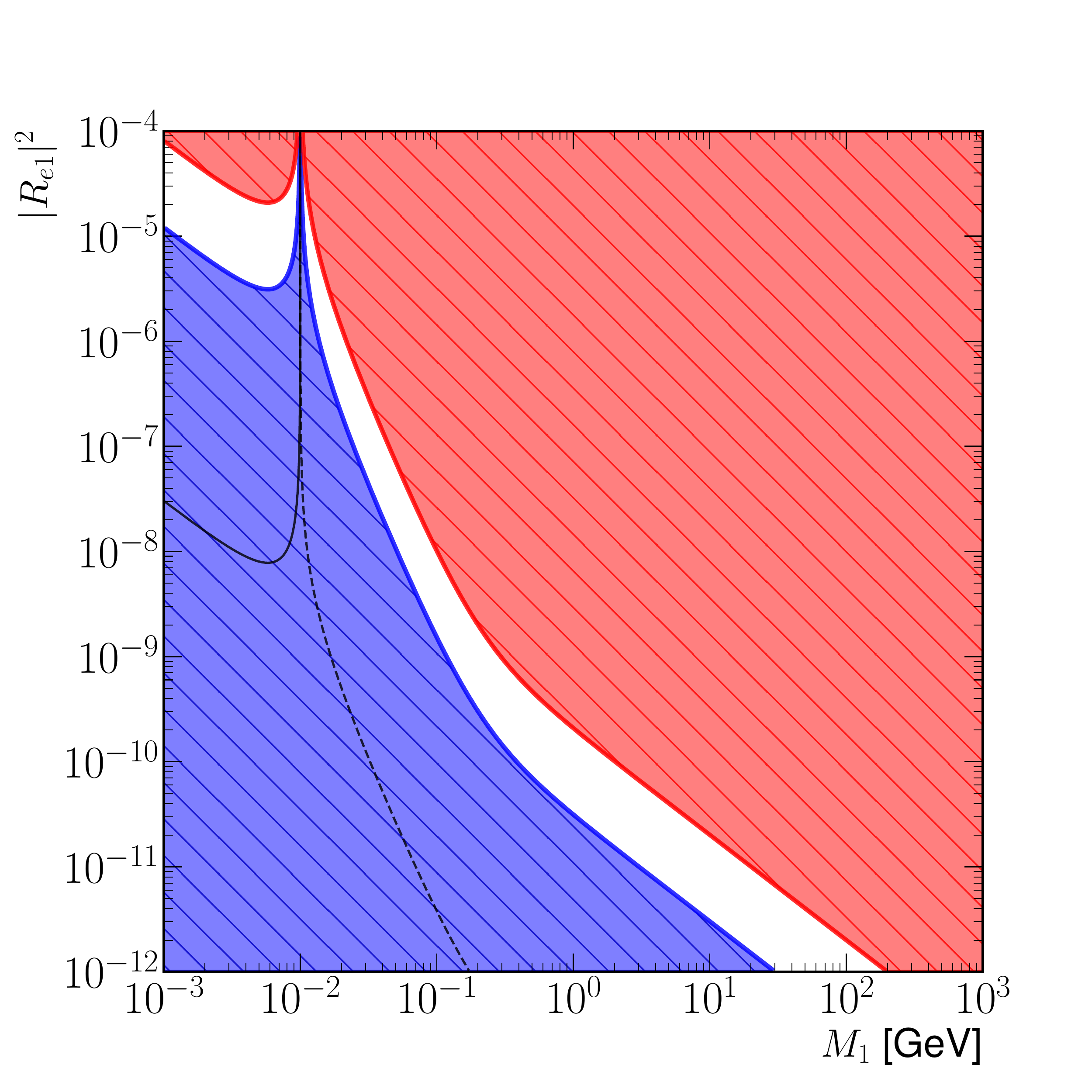}
		}
		\hspace{-0.8cm}
		\subfigure[IH, $M_{2}=200 ~\mathrm{MeV}$]{
			\includegraphics[width=0.34\textwidth]{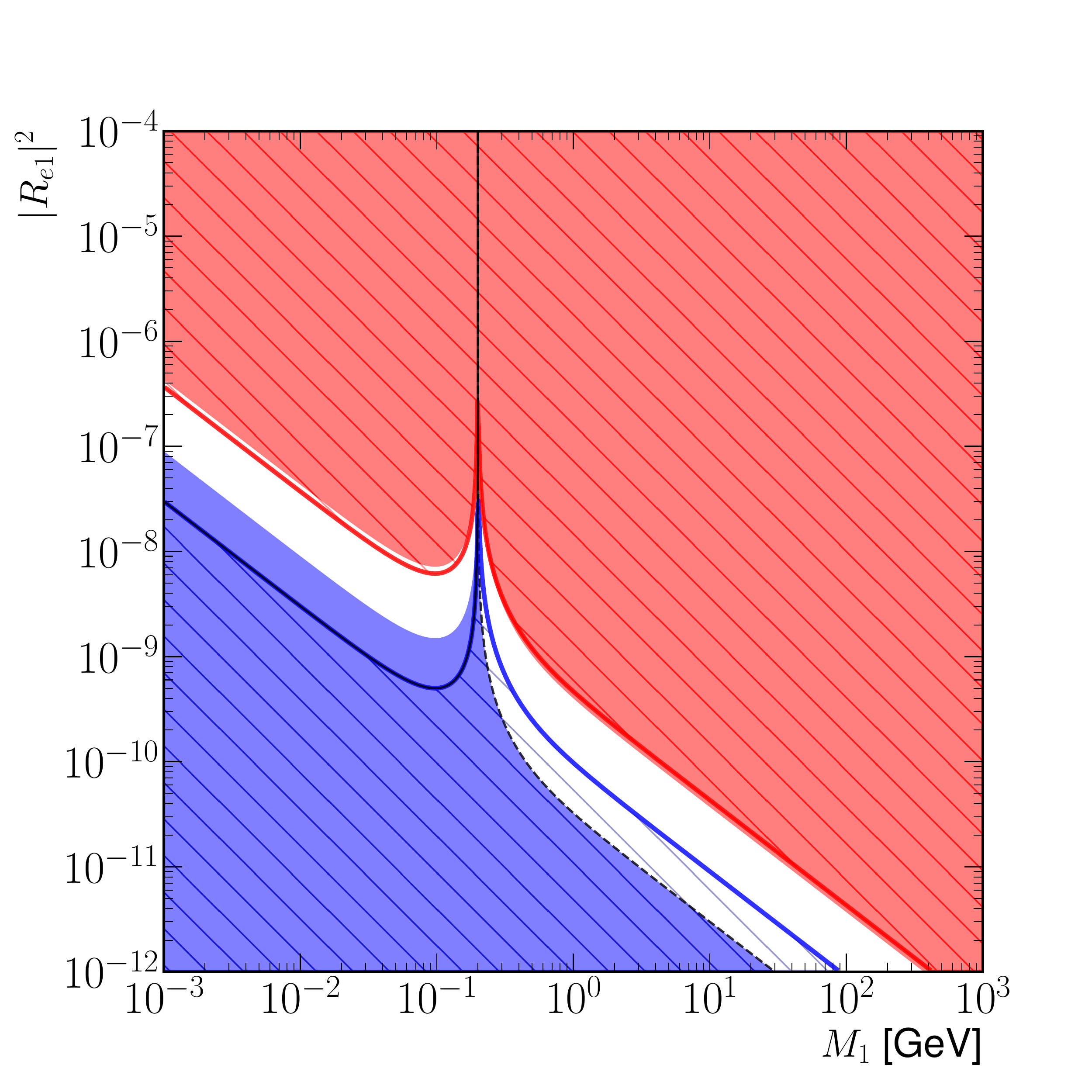}
		}
		\hspace{-0.8cm}
		\subfigure[IH, $M_{2}=1 ~\mathrm{TeV}$]{
			\includegraphics[width=0.34\textwidth]{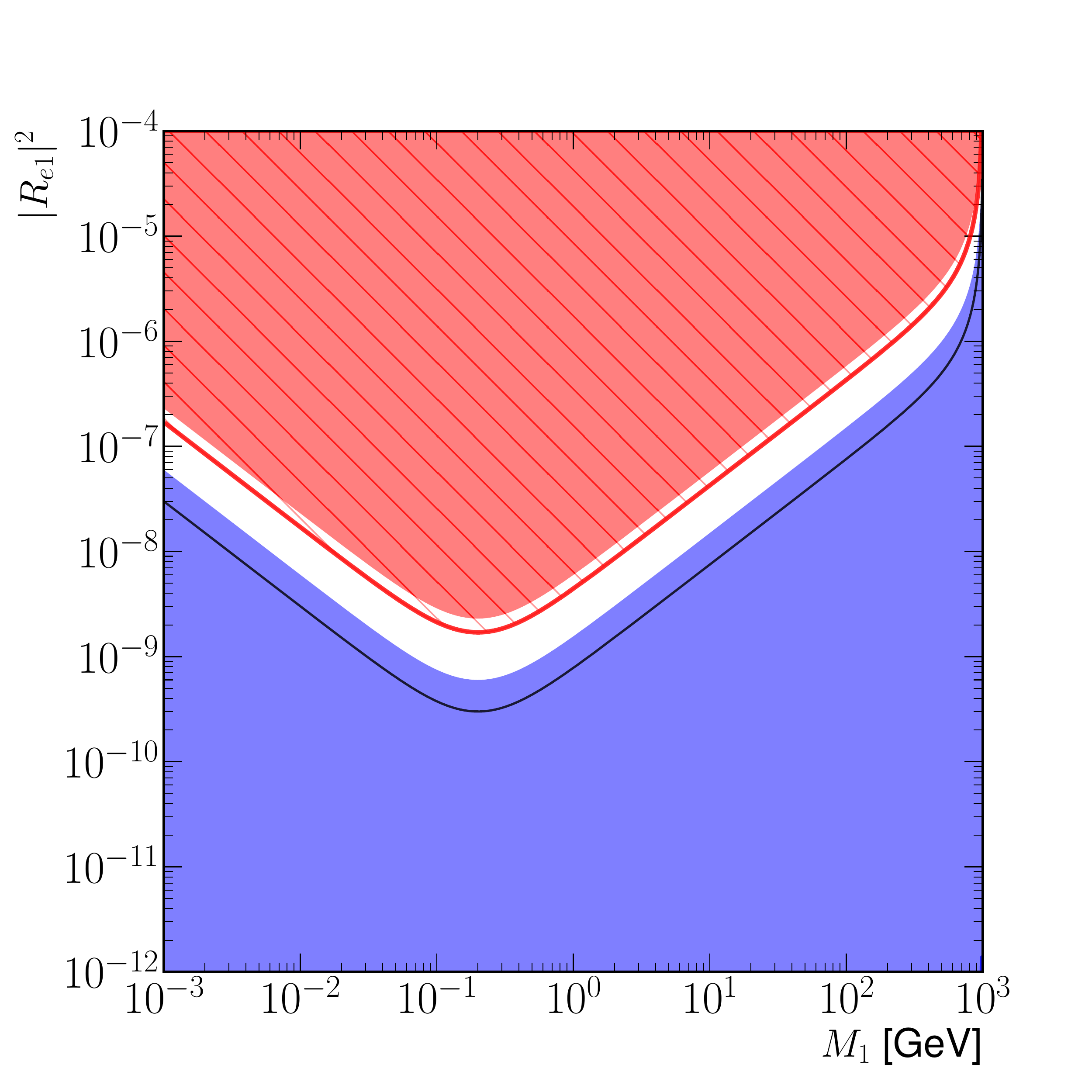}
		}
		\caption{
		Contours of $|m_{\mathrm{eff}}|$ in the $0\nu\beta\beta$-decay as functions of $|R_{e1}^{2}|$ and {$M_{1}$} for the NH (upper panels) and IH (lower panels) cases. 
		We take $\delta_{14}=\pi/2$ (colored regions without slash lines)
		and $\delta_{14}=0$ (regions with slash lines).
		The black solid line and dashed line stand for $m_{\mathrm{eff}}=0$ in $\delta_{14}=\pi/2$ and $\delta_{14}=0$, respectively.
		The blue regions stand for $|m_{\mathrm{eff}}|<|m_{\mathrm{eff}}^{\nu}|$,
		the red regions stand for $|m_{\mathrm{eff}}|>200~\mathrm{meV}$,
		and the white regions correspond to the moderate enhancement in $|m_{\mathrm{eff}}|$.
		}
		\label{fig:5}
	\end{figure}

In order to reveal the quantitative parameter dependence regarding the contours of enhancement and cancellation,
we further cut the contours of Fig.~\ref{fig:3d can} in the horizontal direction by fixing the parameter of $|R_{e1}^2|$ in Fig.~\ref{fig:4} and in the vertical direction by fixing the parameter of $M_2$ in Fig.~\ref{fig:5}.

	In Fig.~\ref{fig:4}, we illustrate the contours of $|m_{\mathrm{eff}}|$ in the left, middle and right panels with the fixed values of $|R_{e1}^2|=10^{-8}, 10^{-10}, 10^{-12}$. The upper and lower panels show the results for the NH and IH cases respectively.
	In each plot, we also take the phase $\delta_{14}=\pi/2$ (colored region without slash lines) and $\delta_{14}=0$ (region with slash lines) respectively.
    The red regions represent the contours of strong enhancement with $|m_{\mathrm{eff}}|>200 ~\mathrm{meV}$ and 
    the blue regions show the cancellation contours with  $|m_{\mathrm{eff}}|<|m_{\mathrm{eff}}^{\nu}|$.
	The special case with the full cancellation, which means $|m_{\mathrm{eff}}|=0$ is also shown on the figures with black lines (solid for $\delta=\pi/2$ and dashed for $\delta=0$).
	The white regions in between are for the case of moderate enhancement allowed by current experiments.
	The strong enhancement always happens in the case with large $M_1$ and small $M_2$, and it is positively correlated to the mixing matrix element $|R_{e1}^2|$ as well as the hierarchy of the two masses of RHNs.
	In the case of large mixing $|R_{e1}^2|$, the regions of strong enhancement also appear around the island at $M_1=0.1 ~\mathrm{GeV}$ with $M_2>0.1 ~\mathrm{GeV}$.
	The boundary between the white and blue regions is shown for $|m_{\mathrm{eff}}|=|m_{\mathrm{eff}}^{\nu}|$, which stands for the case where the contribution of RHNs disappears.
	As we discussed above, the cancellation always happens when both of the RHNs are light, and the cancellation region expands to a larger region when $|R_{e1}^2|$ becomes smaller.
	When the value of $M_1$ is large, the cancellation also happens around the mass degeneracy regions. 
    {In the mass degeneracy case, $|m_{\mathrm{eff}}|$ is always smaller than $|m_{\mathrm{eff}}^{\nu}|$ and the cancellation always happens, but if there is a tiny difference between $M_1$ and $M_2$, the cancellation is rapidly changing to enhancement around $M_1=\sqrt{\langle p^{2}\rangle}$ with large mixing $|R_{e1}^2|$. 
	These special situations can be clearly found in Fig.~\ref{fig:4a} and ~\ref{fig:4d}.
	And of course, it will disappear when $|R_{e1}^2|$ is small enough due to the fact that $|m_{\mathrm{eff}}|$ is smaller than $|m_{\mathrm{eff}}^{\nu}|$ in almost all regions.}
	Comparing the case of $\delta_{14}=0$ with that of $\delta_{14}=\pi/2$, the regions of strong enhancement change slightly, while the regions of cancellation change drastically, covering quite a different parameter space. 
    {The reason is that when the contribution of RHNs is large enough, it will be dominant in $|m_{\mathrm{eff}}|$ so that $|m_{\mathrm{eff}}|$ is not sensitive to the relative phase $\delta_{14}$. 
	On the other hand, it is sensitive when $|m_{\mathrm{eff}}^{N}|$ is close to $|m_{\mathrm{eff}}^{\nu}|$, and in this case, the phase $\delta_{14}$ will determine whether there is enhancement or cancellation.}

Here we want to remark that the intrinsic relation in the seesaw mechanism leads to the removal of the free parameter $R_{e2}^2$, which also breaks the commutative symmetry of $N_1$ and $N_2$. With a given set of $M_1$ ,$M_2$ and $R_{e1}^2$, exchanging the masses of $N_1$ and $N_2$ can not ensure the values of mixing matrix elements are also exchanged. Because of this, the absolute values of $m_{\mathrm{eff}}^{N}$ and regions of cancellation or enhancement are not symmetric, and strong enhancement close to $|m_{\mathrm{eff}}^{\nu}|$ can not happen in case of $\delta_{14}=0$ with small mixing $|R_{e1}^2|$. 
	
	
	\begin{figure}[!htbp]
		\centering
		\subfigure[NH, $M_{1}=10 ~\mathrm{MeV}$]{
			\includegraphics[width=0.4\textwidth]{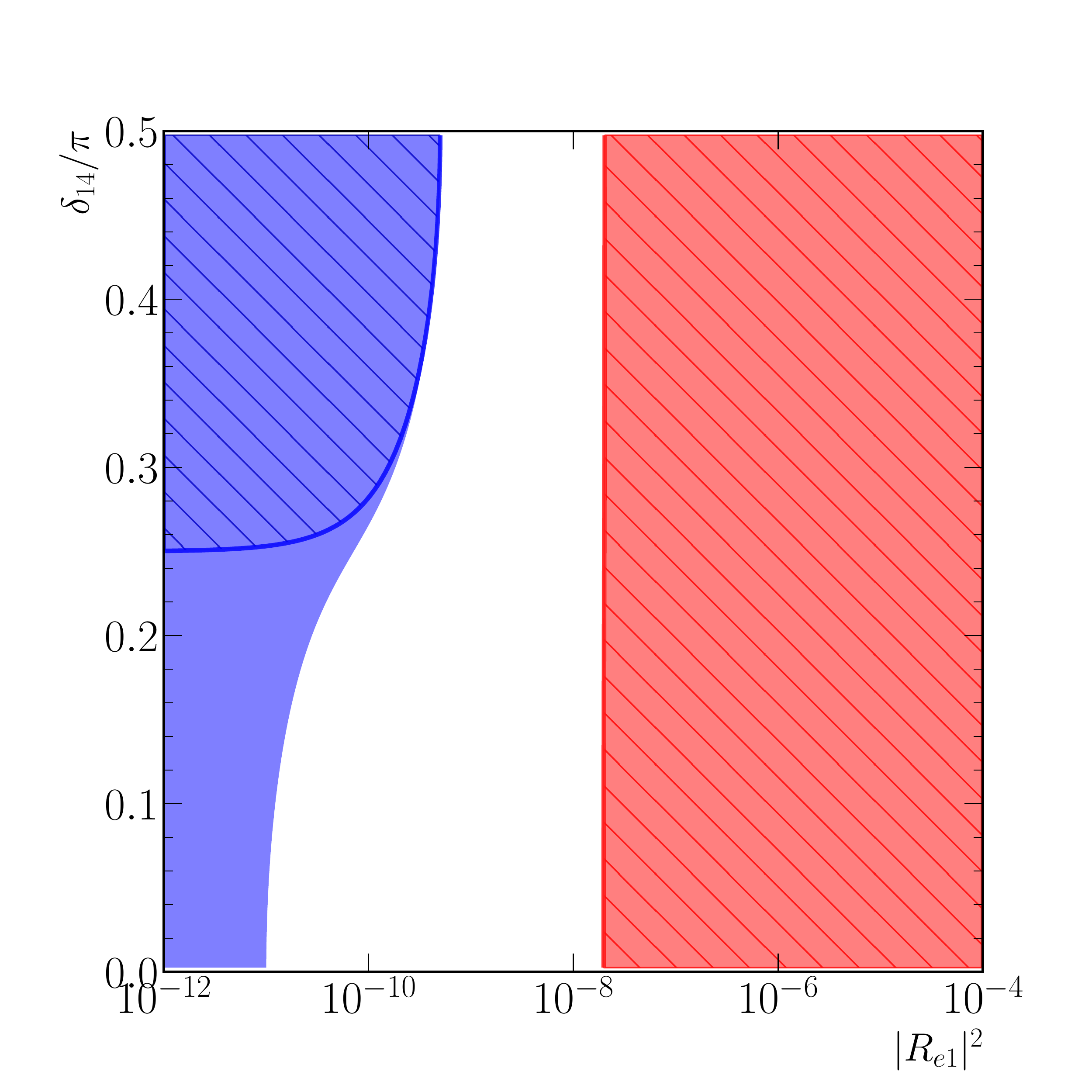}
		}
		\hspace{-0.5cm}
		\subfigure[NH, $M_{1}=200 ~\mathrm{MeV}$]{
			\includegraphics[width=0.4\textwidth]{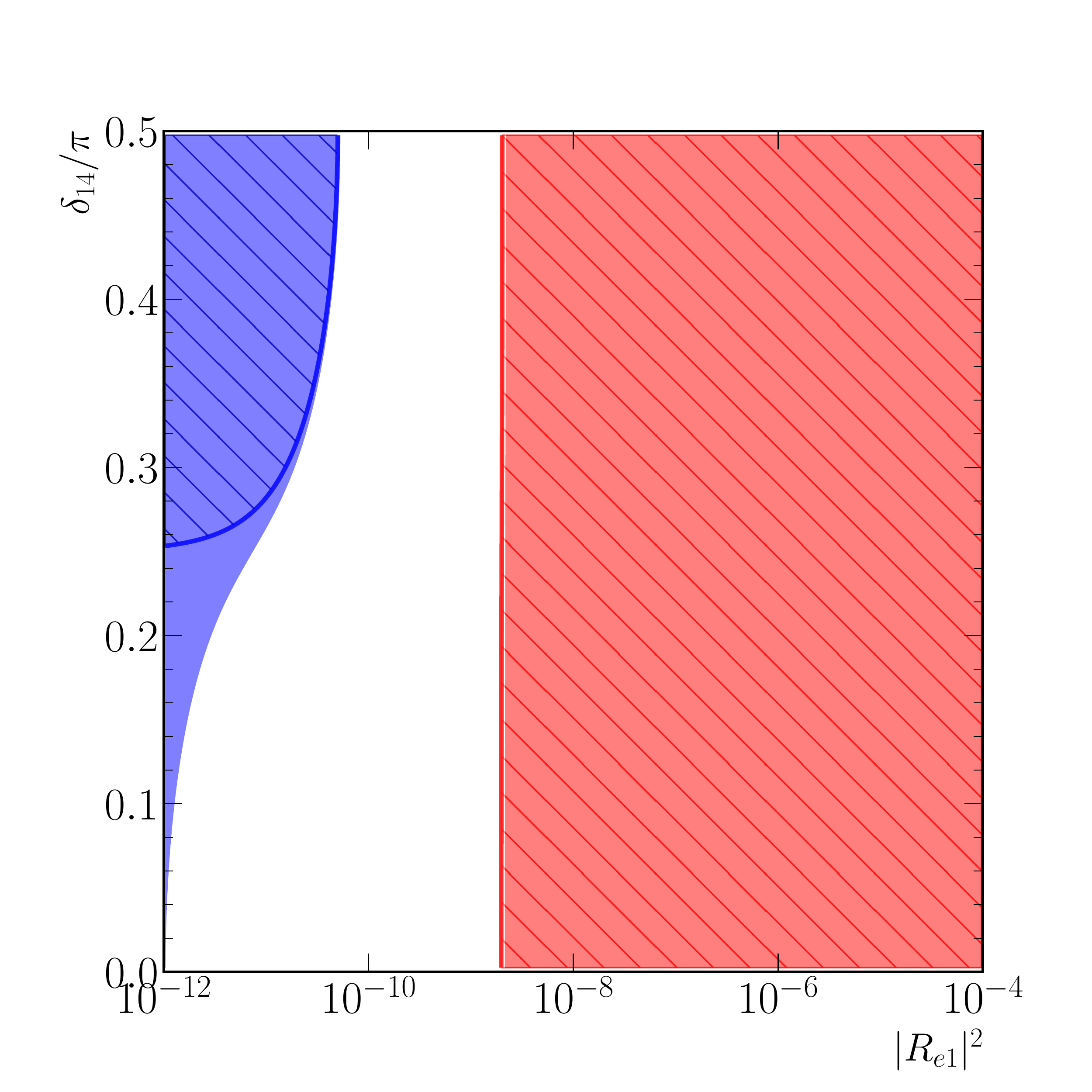}
		}
		
		\subfigure[IH, $M_{1}=10 ~\mathrm{MeV}$]{
			\includegraphics[width=0.4\textwidth]{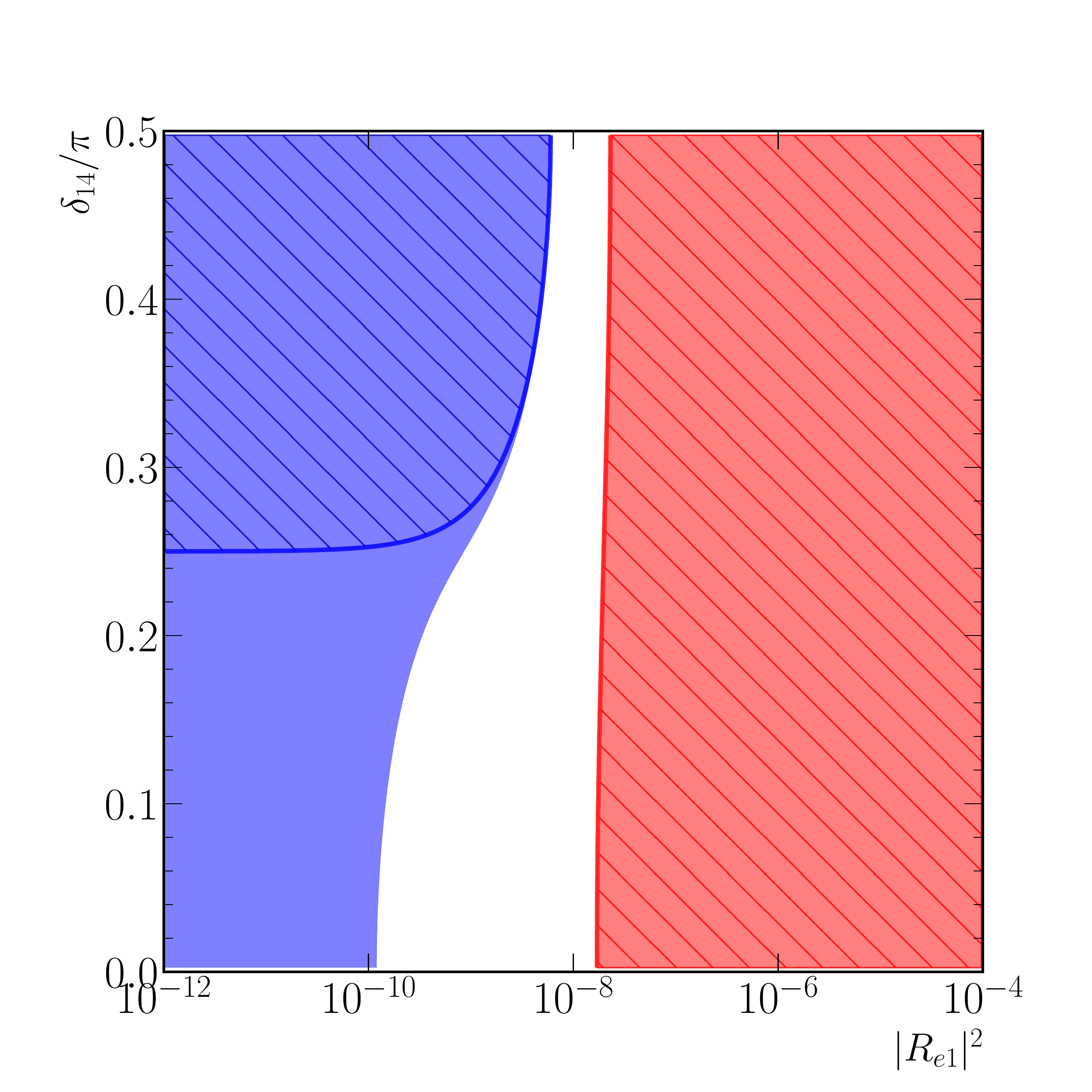}
		}
		\hspace{-0.5cm}
		\subfigure[IH, $M_{1}=200 ~\mathrm{MeV}$]{
			\includegraphics[width=0.4\textwidth]{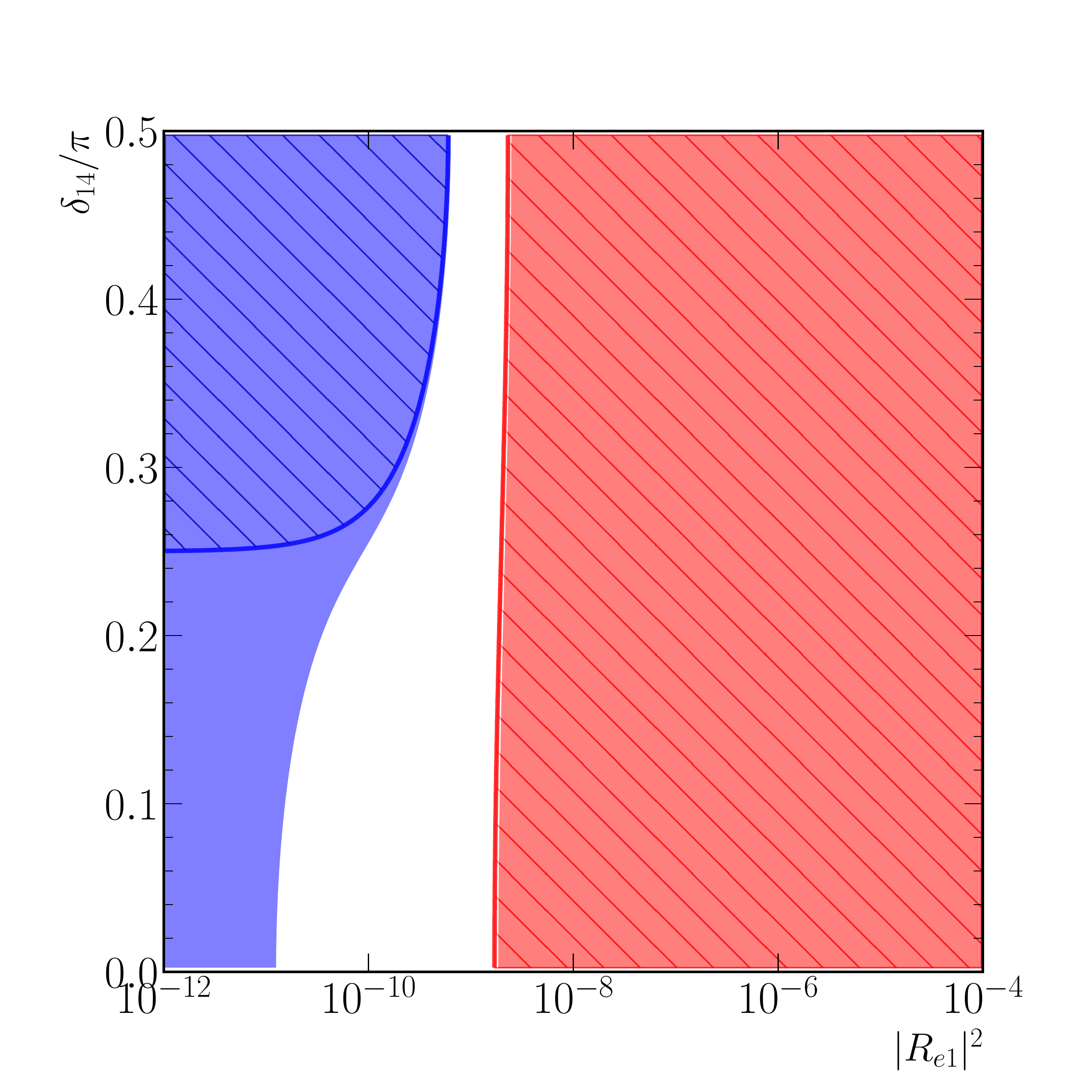}
		}
		\caption{Contours of $|m_{\mathrm{eff}}|$ in the $0\nu\beta\beta$-decay as functions of $|R_{e1}^{2}|$ and {$\delta_{14}$} for the NH (upper panels) and IH (lower panels) cases. 
		We take $M_{1}=10, 200 ~\mathrm{MeV}$ with
		$M_{2}=1 ~\mathrm{GeV}$ (colored regions without slash lines) and $M_{2}=1 ~\mathrm{TeV}$ (regions with slash lines). The blue regions stand for $|m_{\mathrm{eff}}|<|m_{\mathrm{eff}}^{\nu}|$,
	    the red regions stand for $|m_{\mathrm{eff}}|>200~\mathrm{meV}$,
		and the white regions correspond to the moderate enhancement in $|m_{\mathrm{eff}}|$.
		}
		\label{fig:2D R1 a1}
	\end{figure}
	
	Now in Fig.~\ref{fig:5} we discuss the role of the mixing element $|R_{e1}^{2}|$  in the contours of cancellation and enhancement. 
	If future experiments determine a precise effective neutrino mass $m_{\mathrm{eff}}$, the masses of RHNs and the mixing matrix element can be well constrained.
	The values of $|R_{e1}^{2}|$ can be directly determined by a measurement of $|m_{\mathrm{eff}}|$, a given $|m_{\mathrm{eff}}^{\nu}|$ and two masses of RHNs, in which $|m_{\mathrm{eff}}^{\nu}|$ can be properly determined from neutrino oscillation data. 
	In this sense, we can explore the allowed and excluded regions of $|R_{e1}^{2}|$ in Fig.~\ref{fig:5} as functions of a given mass $M_1$ for both NH and IH cases.
	We take the mass of $N_2$ to be $M_2=0.01,0.2,1000 ~\mathrm{GeV}$, and assume the observed values of the effective neutrino mass to 
	$|m_{\mathrm{eff}}|>200~\mathrm{meV}$ (strong enhancement), $|m_{\mathrm{eff}}|<|m_{\mathrm{eff}}^{\nu}|$ (cancellation) or in between. 
	We find that strong enhancement regions are similar in both the cases of NH and IH.
    {When the strong enhancement happens, $|m^{N}_{\mathrm{eff}}|$ is dominant, and the effect of $|m_{\mathrm{eff}}^{\nu}|$ or the phase $\delta_{14}$ can be ignored. In contrast, when the cancellation actually happens, even though the active neutrino mixing and masses could be well determined by neutrino oscillation experiments, 
    the cancellation regions will be drastically changed since the value of $|R_{e1}^{2}|$ is sensitive to the phase $\delta_{14}$.} The mass degeneracy effect is also clearly shown in the figure. When $M_1=M_2$, the value of $|R_{e1}^{2}|$ cannot be constrained by the measurement of $|m_{\mathrm{eff}}|$ since in this case, $|R_{e1}^{2}|$ becomes singular. The black lines stand for $|m_{\mathrm{eff}}|=0$ where the $0\nu\beta\beta$-decay disappears.
    If the two RHNs are with large enough masses, $|R^2_{e1}|$ can be well constrained with a known $|m_{\rm{eff}}|$ indirectly by the intrinsic seesaw relation.
	For $M_2$ around $\langle p^2\rangle$, we also find that the phase will change the cancellation regions, smaller $|R^2_{e1}|$ are needed for the cancellation to happen with larger $\delta_{14}$ in the case of $M_1>M_2$, contrary to the case of $M_1<M_2$. 
	In the case of $M_2=1~\mathrm{TeV}$, the cancellation disappears with $\delta_{14}=0$ for $|R^2_{e1}|$ as large as $10^{-12}$.

	Finally, to get a better understanding on the role of $\delta_{14}$, we show in Fig.~\ref{fig:2D R1 a1} more details of the effect of $\delta_{14}$ on the effective neutrino mass $|m_{\mathrm{eff}}|$.
	We discuss the cases of $M_1=10, 200 ~\mathrm{MeV}$ by taking the values of $M_2$ at $1 ~\mathrm{GeV}$ and $1 ~\mathrm{TeV}$.
	The reason for this consideration is that we could avoid the mass degeneracy by forcing $M_1<M_2$.
	We find that the cases of $\delta_{14}=0$ and $\delta_{14}=\pi/2$ in the previous figures present two extreme cases, all the other choices of $\delta_{14}$ are in between these two sets of results. Therefore, choosing these two values generally gives a full view of the $|m_{\rm{eff}}|$ on the phase.
	For the cases with $M_1<M_2$, we find that the strong enhancement regions are basically independent of $\delta_{14}$, as already noted in the lower-right parts of Fig.~\ref{fig:3d can}. The cancellation happens for smaller $|R^2_{e1}|$ if we decrease $\delta_{14}$. The mass difference of the two RHNs will affect the behaviour of $m_{\rm{eff}}$ as stated before. We observe that the cancellation is very sensitive not only to the absolute masses of the RHNs but also their mass difference. A larger mass difference leads to weaker cancellation, in particular for smaller values of $\delta_{14}$.
	
	Before finishing this section, we would like to summarize the experimental constraints on the parameter space of RHNs from other probes and discuss the rationale of our choices of the typical parameter range. Firstly, assuming the electron neutrino mixing dominance, the mixing $|R^2_{e1}|$ is constrained to the level of [$10^{-8}$, $10^{-9}$] by NA62~\cite{NA62:2020mcv} for the $M_1$ range between 150 MeV and 400 MeV, and the level of [$10^{-6}$, $10^{-7}$] by CHARM~\cite{CHARM:1985nku} for $400\;{\rm MeV}<M_1<2\;{\rm GeV}$. Constraints from Belle~\cite{Belle:2013ytx}, DELPHI~\cite{DELPHI:1996qcc}, ATLAS~\cite{ATLAS:2019kpx}, CMS~\cite{CMS:2018iaf} can extend to the region of $M_1>2\;{\rm GeV}$, which present a limit of $|R^2_{e1}|$ at the level of $10^{-4}$.
	Secondly, regarding the neutrino probe, T2K~\cite{T2K:2019jwa} is sensitive to $|R^2_{e1}|$ for the mass range of $150\;{\rm MeV}<M_1<500\;{\rm MeV}$, in which the limit varies from $10^{-7}$ to $10^{-9}$. Meanwhile MicroBooNE~\cite{MicroBooNE:2019izn} and ArgoNeuT~\cite{ArgoNeuT:2021clc} have also made the searches of RHNs at the same mass range, but their constraints are for the muon and tau neutrino channels respectively. Thirdly, the process of charged lepton flavor violation is sensitive to the combination of $|\Sigma_{i} R^{}_{\alpha i}R^{*}_{\beta i}|$~\cite{Lindner:2016bgg}, where $\alpha$ and $\beta$ are the final and initial flavors of the process, respectively, and $i$ is the index of the RHNs. The present bound on $|\Sigma_{i} R^{}_{\alpha i}R^{*}_{\beta i}|$ is at the level of [$10^{-3}$, $10^{-4}$] for the GeV-scale RHNs~\cite{Alonso:2012ji}.
	However, the above experimental searches of RHNs are typically carried out under the assumption of a single RHN mixing with a single flavor, which cannot be used in the realistic seesaw model.
	A recent study has shown that the limit on the mixing element could be several orders weaker than the experimental reported ones after marginalizing over the unknown model parameters~\cite{Tastet:2021vwp}. Therefore, our choices of $|R^2_{e1}|$ are motivated by the above experimental limits and the largest one of $10^{-8}$ can be considered near the edge of the current upper bound, which is also motivated by the seesaw relation where strong cancellation between two RHNs is required to produce the correct active neutrino masses. 
	
	\section{Conclusion}
	\label{sec:conclusion}
	
	In this work, we have studied the contribution to the effective neutrino mass of the $0\nu\beta\beta$-decay from two RHNs within the minimal type-I seesaw model using the intrinsic seesaw relation of neutrino mass and mixing parameters and the relative mass dependence of NMEs. The effective neutrino mass as a function of the parameters of the RHNs is illustrated. We have demonstrated that both the enhancement and cancellation to the effective neutrino mass from the RHNs can happen, with a quantitative exploration on their dependence on the masses $M_{1}$, $M_{2}$, the mixing element $|R^2_{e1}|$ and the phase $\delta_{14}$. 
	In general, strong enhancement usually needs the large mass hierarchy of two RHNs, but it also happens for the masses of the RHNs at around $ 0.2~\mathrm{GeV}$ with large mixing element $|R^2_{e1}|$.
	The cancellation happens in the case that the masses of the RHNs and the mixing element $|R^2_{e1}|$ are relatively small, but there are also exception for large masses in the presence of strong cancellation between the RHNs themselves, in which the phase plays a crucial role.
	We believe our studies in this work can be useful to analyze the current and future experimental results and help to pin down the viable parameter space of the seesaw model. We defer this to future separated works. 
	
	
	\begin{acknowledgements}
		The work of  Y.F. Li and Y.Y. Zhang is supported by the National Natural Science Foundation of China under Grant No.~12075255 and No.~11835013,
		and by Beijing Natural Science Foundation under Grant No.~1192019.
		Y.F. Li is also grateful for the support by the CAS Center for Excellence in Particle Physics (CCEPP).
		The work of Y.Y. Zhang is also supported by China Postdoctoral Science Foundation under Grant No.~2021T140669.
		DLF is supported by CAS from the "Light of West" program and the "from zero to one" program.
	\end{acknowledgements}

%

\end{document}